\documentclass[preprint,12pt]{elsarticle}
\usepackage[margin=2.5cm]{geometry}
\usepackage{lineno}
\usepackage{setspace}

\RequirePackage{amsthm,amsmath,amsfonts,amssymb}
\RequirePackage[numbers]{natbib}
\RequirePackage[colorlinks,citecolor=blue,urlcolor=blue,backref=page,backref=page]{hyperref}
\RequirePackage{graphicx}

\usepackage{psfrag}
\usepackage{subfigure}
\usepackage{amssymb} 

\usepackage{algorithmic}

\newcommand{\convL}{\overset{\mathcal{L}}{\rightarrow}}

\usepackage[ruled,vlined]{algorithm2e}

\newcommand{\R}{\mathbb{R}}

\newcommand{\g}[1]{\boldsymbol{#1}}
\newcommand{\PP}[1]{\left( #1 \right)}

\newcommand{\Ac}[1]{\left\{ #1 \right\}}
\newcommand{\Cr}[1]{\left[ #1 \right]}

\newcommand{\Eq}[1]{\begin{equation} #1 \end{equation}}
\newcommand{\Eqa}[1]{\begin{equation*} #1 \end{equation*}}

\newcommand{\NOR}[1]{\left\Vert  #1 \right\Vert}

\newcommand{\BI}[1]{\begin{itemize} #1 \end{itemize}}

\newcommand{\BFig}[1]{\begin{figure}\begin{center} #1 \end{center}\end{figure}}

\newcommand{\comm}[1]{}

\graphicspath{{images/}}

\makeatletter
\def\blfootnote{\gdef\@thefnmark{}\@footnotetext}
\makeatother

\newcommand{\COR}[1]{{\color{blue} }}

\begin{document}
\begin{frontmatter}

\title{Estimating Intractable Posterior Distributions through Gaussian Process regression and Metropolis-adjusted Langevin procedure}

\author[label1]{Guillaume Perrin}
\ead{guillaume.perrin@univ-eiffel.fr}

\author[label1,label2,label3]{Romain Jorge Do Marco}
\ead{romain.jorge-domarco@sncf.fr}

\author[label2]{Christian Soize}
\ead{christian.soize@univ-eiffel.fr}

\author[label3]{Christine Funfschilling}
\ead{christine.funfschilling@sncf.fr}

\address[label1]{Université Gustave Eiffel, COSYS, 14-20 Boulevard Newton, 77447 Marne-la-Vallée, France}
\address[label2]{Université Gustave Eiffel, {MSME UMR 8208 CNRS}, {5 bd Descartes}, {Marne-la-Vall\'{e}e}, {77454}, {France}}
\address[label3]{SNCF, DTIPG, 1-3 Av François Mitterrand, 93574 Saint-Denis, France}

%
%

\begin{abstract}
Numerical simulations are crucial for modeling complex systems, but calibrating them becomes challenging when data are noisy or incomplete and likelihood evaluations are computationally expensive. Bayesian calibration offers an interesting way to handle uncertainty, yet computing the posterior distribution remains a major challenge under such conditions. To address this, we propose a sequential surrogate-based approach that incrementally improves the approximation of the log-likelihood using Gaussian Process Regression. Starting from limited evaluations, the surrogate and its gradient are refined step by step. At each iteration, new evaluations of the expensive likelihood are added only at informative locations, that is to say where the surrogate is most uncertain and where the potential impact on the posterior is greatest. 
The surrogate is then coupled with the Metropolis-Adjusted Langevin Algorithm, which uses gradient information to efficiently explore the posterior. This approach accelerates convergence, handles relatively high-dimensional settings, and keeps computational costs low. We demonstrate its effectiveness on both a synthetic benchmark and an industrial application involving the calibration of high-speed train parameters from incomplete sensor data.
\end{abstract}

\begin{keyword}
{Bayesian inference};
{statistical learning};
{Markov Chain Monte Carlo};
{filtering}.
\end{keyword}

\end{frontmatter}

\section{Introduction}
\label{sec1}

In many fields of engineering and physics, numerical simulation is a fundamental tool for modeling and predicting the behavior of complex systems. These models, often based on differential equations or phenomenological representations, require the estimation of parameters to ensure their predictive accuracy. Such parameters may include physical constants, numerical thresholds or other critical values that describe the behavior of the system under different conditions. For the sake of clarity, 
we call environment parameters the parameters that allow us to characterize the system environment (such as temperature, pressure, boundary conditions, etc.), and we call calibration parameters these parameters that need to be specified for the code to be run, and which are assumed not to depend on the environmental conditions. Unsurprisingly, the accuracy and reliability of a simulation are highly sensitive to the correct tuning of the calibration parameters. Consequently, the accurate estimation of these parameters in the presence of noisy and incomplete data is a major challenge with wide implications in scientific research and industry.
The Bayesian framework, particularly as formulated by Kennedy and O'Hagan \cite{kennedy2001bayesian}, has become a popular approach to address these challenges. This framework allows the formal incorporation of prior knowledge about calibration parameters, together with the uncertainties inherent in both the measurements and the environment, to estimate the posterior distribution of these calibration parameters given the observed data, taking into account all sources of variability and uncertainty \cite{Oakley2002,Marin2007}.
In practice, however, the computation of this posterior distribution is highly non-trivial. The likelihood functions involved are often computationally expensive to evaluate, as they generally depend on running complex numerical codes. Furthermore, the need to integrate over uncertain environmental variables adds an additional layer of difficulty, leading to what is generally called \textit{intractable} posterior distributions. This intractability arises because evaluating these distributions requires integration over high-dimensional and often poorly understood spaces for which analytical solutions are not feasible, and for which traditional Monte Carlo methods, while theoretically sound, become inefficient or impractical  (see \cite{berger1999integrated} for a discussion of the difficulties associated with these integrated likelihood function methods).
Our approach to this problem is in line with methods seeking to couple advanced machine learning and computational statistics techniques to build relevant approximations of this posterior at a reduced computational cost (see \cite{llorente2024survey} for a review), with the particularity of seeking to associate an error with this approximation. Specifically, we propose to use Gaussian Process Regression (GPR) to model and approximate the expensive log-likelihood function, as also proposed in \cite{jarvenpaa2021parallel} and \cite{jarvenpaa2024approximate}. GPR is a powerful method that can quantify uncertainty and help reduce the number of costly simulations required \cite{Sacks1989,Santner2003,gramacy2020surrogates}. By learning the log-likelihood function, we can efficiently explore the parameter space and make informed decisions about where to evaluate the expensive numerical code next.
In addition, we integrate this GPR with the Metropolis-Adjusted Langevin Algorithm (MALA), a Markov Chain Monte Carlo (MCMC) method. MALA is particularly well suited to high-dimensional and complex distributions because it uses gradient information to propose efficient updates, thereby improving convergence speed \cite{roberts1996exponential,roberts1998optimal}. By combining the GPR surrogate model with MALA, we aim to construct an efficient strategy for posterior inference: the GPR model provides an approximation of the log-likelihood function and its gradient (as it is done in \cite{perrin2021constrained} and \cite{perrin2024bayesian}), while MALA ensures that the exploration of the posterior distribution is robust and accurate. 
Overall, our approach seeks to balance accuracy and computational efficiency in the estimation of these intractable posterior distributions. In doing so, we aim to make Bayesian inference of complex numerical models more feasible, providing a framework that can be applied to a wide range of scientific and engineering problems where data-driven model inference is essential. 
Even if they do not integrate this information on the GPR gradient, it should be noted that other works in the literature have also proposed to couple GPR and MCMC approaches. This is for instance the case in \cite{jarvenpaa2024approximate}, which seeks to integrate the uncertainty linked to the GPR into the acceptance ratio of the MCMC algorithm. In substance, the approach developed in this work nevertheless strongly differs from these works by the fact that it proposes to work trajectory by trajectory of the GPR, in particular by independently repeating the MALA procedures for a better monitoring of the approximation errors. 

The outline of this work is as follows: Section \ref{sec2} presents the theoretical framework of the proposed method, and Section \ref{sec3} illustrates the benefits of such a coupling, firstly, on an analytical case for which all model parameters can be controlled, and then, secondly, on an industrial case for estimating mechanical and energy parameters of a high-speed train from on-board measurements. 

\section{Theoretical framework}
\label{sec2}

\subsection{Problem definition}
\label{sec21}

For $d\geq 1$, let $\mathbb{X}$ be a subset of $\R^d$, $\mathcal{C}^1(\mathbb{X},\R^+)$ be the set of positive functions that are continuous and one time differentiable at all $\g{x}\in\mathbb{X}$, and $\mathbb{H}_1(\mathbb{X},\R^+)$ be the  set of functions that are in $\mathcal{C}^1(\mathbb{X},\R^+)$ and whose integral over $\mathbb{X}$ is equal to one:


\Eq{\mathbb{H}_1(\mathbb{X},\R^+):=\Ac{ h\in \mathcal{C}^1(\mathbb{X},\R^+): \int_{\mathbb{X}}h(\g{x})d\g{x}=1 }.}

In this work, we are interested in estimating a particular function $h^\star \in \mathbb{H}_1(\mathbb{X},\R^+)$ that can be written as
 
\Eq{ h^\star(\g{x}) = c\times \mathbb{E}_{\g{Z}}\Cr{g(\g{x},\g{Z})}, \quad \g{x}\in\mathbb{X},\label{pb_mini}}

\noindent{}where $c$ is an unknown positive normalization constant, $\g{Z}$ is a random vector with values in $\mathbb{Z}\subset \R^q$, and $g$ is a positive time-consuming deterministic black-box function, in the sense that it returns twice the same value when evaluated twice at the same inputs and that we only have access to pointwise evaluations of $g$ (no knowledge of the phenomena leading to its values or access to its derivatives). 
As explained in Introduction, a typical situation where this kind of problem arises is when we try to infer the parameters of a numerical code in configurations that are not completely controlled. Indeed, such numerical codes are generally based on two kinds of inputs necessary for their execution. First, the system parameters characterize the conditions of the experiment, such as temperature, pressure, system dimensions, boundary conditions, and so on. Secondly, the calibration parameters refer to constants considered universal, meaning they are supposed to be independent of the system's characterizing parameters and usually correspond to physical laws or numerical thresholds (such as laws of crack propagation, turbulence, thermal conductivity, for instance). To make the codes predictive, these calibration parameters must be carefully adjusted based on measurements representative of the system's operating conditions. We can then denote by $\g{x}$ these calibration parameters, by $\g{m}$ the vector of measured data, and by $\g{z}$ the vector containing the environment parameters. If the Bayesian framework of Kennedy and O’Hagan \cite{kennedy2001bayesian} is applied, the calibration parameters are modeled as random variables, to which we assume we can associate an \textit{a priori} probability density function (PDF) noted $f_{\g{X}}$ (using expert knowledge for instance, see \cite{Marin2007} for more information about expert elicitation). In most cases, the experimental measurements are affected by noise, and the environmental parameters are imperfectly known (we can imagine that they are known up to measurement uncertainty, for example). The vectors $\g{z}$ and $\g{m}$ can therefore be modeled by random vectors noted $\g{Z}$ and $\g{M}$ respectively.  Estimating the value of the calibration parameters in such a Bayesian formalism then comes down to estimating the probability distribution of $\g{X}$ conditional on the available observation $\g{m}$ of $\g{M}$, which, using Bayes' formula, can be written as

\Eq{ f_{\g{X}\vert \g{M}=\g{m}}(\g{x}) = \frac{f_{\g{M}\vert \g{X}=\g{x}}(\g{m})  f_{\g{X}}(\g{x})}{ \int_{\mathbb{X}} f_{\g{M}\vert \g{X}=\g{x}'}(\g{m})  f_{\g{X}}(\g{x}') d\g{x}'}, } 

\noindent{}where $f_{\g{M}\vert \g{X}=\g{x}}(\g{m})$ is the likelihood function. By construction, the likelihood function at $\g{x}$ will be greater the closer the numerical code outputs are to the measurements $\g{m}$. However, in order to be evaluated, the numerical code needs to specify not only the values of $\g{x}$, but also the values of the environment variables. To integrate this dependence, we can rewrite the likelihood function in the form:

\Eq{f_{\g{M}\vert \g{X}=\g{x}}(\g{m}) = \int_{\mathbb{Z}} f_{\g{M},\g{Z}\vert \g{X}=\g{x}}(\g{m},\g{z}) d\g{z},}

\noindent{}with $\mathbb{Z}$ the definition domain of the random vector $\g{Z}$ gathering the possible values of the environment variables. Then, assuming that the uncertainties affecting the values of $\g{Z}$ are {statistically} independent of the value of $\g{X}$, that is to say assuming that $f_{\g{Z}\vert \g{X}=\g{x}}(\g{z})=f_{\g{Z}}(\g{z})$, and using the Bayes formula a second time, we deduce:

\Eq{f_{\g{M}\vert \g{X}=\g{x}}(\g{m}) \ \propto \ \int_{\mathbb{Z}} f_{\g{M}\vert \g{X}=\g{x},\g{Z}=\g{z}}(\g{m}) f_{\g{Z}}(\g{z}) d\g{z}=: \mathbb{E}_{\g{Z}}\Cr{ f_{\g{M}\vert \g{X}=\g{x},\g{Z}}(\g{m}) },}

\noindent{}where the symbol $\propto$ indicates a proportional relationship to a multiplicative constant. It finally comes:

\Eq{f_{\g{X}\vert \g{M}=\g{m}}(\g{x}) = \frac{  \mathbb{E}_{\g{Z}}\Cr{ f_{\g{M}\vert \g{X}=\g{x},\g{Z}}(\g{m}) } f_{\g{X}}(\g{x})}{ \int_{\mathbb{X}} \mathbb{E}_{\g{Z}}\Cr{ f_{\g{M}\vert \g{X}=\g{x}',\g{Z}}(\g{m}) } f_{\g{X}}(\g{x}') d\g{x}'}, \quad \g{x}\in\mathbb{X}.}

Now, if we write $g(\g{x},\g{z})= f_{\g{M}\vert \g{X}=\g{x},\g{Z}=\g{z}}(\g{m}) f_{\g{X}}(\g{x})$ for each $(\g{x},\g{z})$ in $\mathbb{X}\times \mathbb{Z} $,  and $c^{-1}=\int_{\mathbb{X}} \mathbb{E}_{\g{Z}}\Cr{ f_{\g{M}\vert \g{X}=\g{x}',\g{Z}}(\g{m}) } f_{\g{X}}(\g{x}') d\g{x}'$, we find back the formalism of Eq. (\ref{pb_mini}), the function $h^\star$ corresponding to the \textit{a posteriori} PDF of $\g{X}$ knowing the observation $\g{M}=\g{m}$.

\subsection{Specificities, difficulties and proposed methodology}
\label{sec22}

The approximation of the function $h^\star$ raises several important difficulties, for which we propose specific developments, which we list in the remainder of this section.

$\bullet$ \textbf{Presence of an unknown multiplicative constant.}
Two first difficulties arise from the unknown character of $c$, and from the fact that, for each value of $\g{x}$, an expectation must be calculated to estimate $h^\star(\g{x})$. This kind of difficulty naturally leads us to the problems of \emph{intractable likelihoods}, which have already been well studied in the literature. For example, you can refer to \cite{gelman1995bayesian,Rubinstein2008,murphy2012machine,marin2012approximate} for a detailed presentation of methods for dealing with inaccessible likelihoods, via Markov Chain Monte Carlo (MCMC), Variational Inference (VI) or Approximate Bayesian Computation (ABC) approaches. 
In particular,
if the numerical cost of $g$ is negligible, and it is possible to evaluate it in a very large number of couples $(\g{x},\g{z})$ in $\mathbb{X}\times \mathbb{Z}$, a standard way of overcoming these two difficulties would be to abruptly replace the expectation by its Monte Carlo estimator using a very large number of realizations of $\g{Z}$, and then use any MCMC algorithm to draw samples from $h^\star$ without needing to know the value of $c$. These samples could finally be post-processed from statistical non-parametric approaches \cite{perrinSIAM2012,perrinCSDA2017,soize2020sampling,soize2022probabilistic} to reconstruct $h^\star$.

$\bullet$ \textbf{Non negligible computational cost.}
In this work, however, we assume that the numerical cost of $g$ is not negligible (from a few seconds to a few minutes for each evaluation for instance). In this case, MCMC approaches, which are notoriously computationally intensive, cannot be directly applied. On the contrary, it seems necessary to base the MCMC algorithm entirely or partially on a mathematical approximation of $g$, also called surrogate model. 
In our case, there are two options for this approximation. On the one hand, we can try to approximate $(\g{x},\g{z})\mapsto g(\g{x},\g{z})$. In this case, we are working in the augmented space $\mathbb{X}\times \mathbb{Z}$, and we will need to be able to efficiently calculate expectations of this approximation with respect to $\g{Z}$ in order to use it in the MCMC. Depending on the approximation considered and the distribution of $\g{Z}$, such an expectation may be explicit. This is the case, for example, when the function $g$ is approximated by a polynomial function of $\g{x}$ and $\g{z}$, and the components of $\g{Z}$ are independent and Gaussian. On the other hand, we can try to directly approximate $\g{x} \mapsto \mathbb{E}_{\g{Z}}\Cr{g(\g{x},\g{Z})}$. This time, we work in the smaller space $\mathbb{X}$, and the approximation can be used directly in the MCMC algorithm, but we need to be able to estimate $\mathbb{E}_{\g{Z}}\Cr{g(\g{x},\g{Z})}$ for several values of $\g{x}$ to construct this approximation. If this expectation is approximated by a Monte Carlo type sampling method, as is generally the case, it is also important to note that the information available for this approximation is necessarily noisy, requiring the use of approximation techniques capable of handling this noise.

$\bullet$ \textbf{Gaussian process regression.}
Motivated by the inference application with non-perfectly controlled conditions presented at the end of Section \ref{sec21}, for which the dimension of $\g{Z}$ is potentially very large, thus limiting the possibility of learning the dependency structure between $g(\g{x},\g{z})$ and $\g{z}$ at reduced cost,
we restrict ourselves in the remainder of this work to the second configuration introduced, and we concentrate on approximating $\g{x} \mapsto \mathbb{E}_{\g{Z}}\Cr{g(\g{x},\g{Z})}$. 
Assuming that we have access to a cluster of $R$ cores that can be run in parallel, we also limit ourselves to the case where, for each $\g{x}$ in $\mathbb{X}$, the number of independent and identically distributed (i.i.d.) realizations of $\g{Z}$ used to approximate $\mathbb{E}_{\g{Z}}\Cr{g(\g{x},\g{Z})}$ is a multiple of $R$, and we leave as working perspective the extension to more general configurations. To approximate the relationship between $\g{x}$ and $\mathbb{E}_{\g{Z}}\Cr{g(\g{x},\g{Z})}$, we now assume that function $g$ has been evaluated $R$ times in $N$ values of $\mathbb{X}$. The total budget for this approximation is thus equal to $N\times R$, and we denote by $\g{x}_1,\ldots,\g{x}_N$ the considered values of $\g{x}$, by $\g{z}_{n,r}$, with $1\leq r\leq R$ and $1\leq n\leq N$, the r$^{\text{th}}$ realization of $\g{Z}$ associated with the n$^{\text{th}}$ value of $\g{x}$, and by $g_{n,r}:=g(\g{x}_n,\g{z}_{n,r})$ the associated value of $g$. Note that limiting the number of calls to $g$ to $R$ does not prevent us from performing the $R$ evaluations of $g$ twice at the same value of $\g{x}$, as long as different realizations of $\g{Z}$ are considered. In particular, this can be used to obtain expectation approximations of varying precision. And we gather all these available input-output pairs into the set $\mathcal{S}_N$, such that:

\Eq{\mathcal{S}_N:=\Ac{ (\g{x}_1,\g{z}_{1,1},g_{1,1}), \ldots, (\g{x}_N,\g{z}_{N,R},g_{N,R})},}

\noindent{}where it is reminded that $\g{z}_{1,1},\ldots,\g{z}_{N,R}$ are $N\times R$ i.i.d. realizations of $\g{Z}$. To approximate $\g{x} \mapsto \mathbb{E}_{\g{Z}}\Cr{g(\g{x},\g{Z})}$ given $\mathcal{S}_N$, we propose to rely on Gaussian process regression (GPR), which is a commonly used class of surrogate models, as it offers closed-form expressions and an estimation of the prediction error \cite{Sacks1989,Santner2003}. The principle of GPR surrogate modeling is to represent a deterministic function by a conditioned Gaussian process. As long as $\mathbb{E}_{\g{Z}}\Cr{g(\g{x},\g{Z})}$ is positive for all $\g{x}$, we have two options: either model $\g{x} \mapsto \mathbb{E}_{\g{Z}}\Cr{g(\g{x},\g{Z})}$ directly as a Gaussian process, incorporating a positivity constraint as proposed in 
\cite{VEIGA2020106732,perrin2021constrained}, or apply a transformation to this function before modeling it as a Gaussian process. This is the second option we consider in the following, and applying a log transform, we assume that $\g{x}\mapsto y(\g{x}):=\log\PP{\mathbb{E}_{\g{Z}}\Cr{g(\g{x},\g{Z})}}$ is a particular realization of a Gaussian process $Y\sim\text{GP}(\mu,C)$, whose mean and covariance functions are denoted by $\mu$ and $C$ respectively. Note that this assumption amounts to supposing that $h^\star$ is a particular realization of the positive random field $H:=c\times \exp(Y)$.
Using $\mathcal{S}_N$, let us now write $\g{y}:=(y_1,\ldots,y_N)$, with

\Eq{y_n:=\log\PP{g_n}, \quad g_n:=\frac{1}{R}\sum_{r=1}^R g(\g{x}_n,\g{z}_{n,r}), \ \ 1\leq n\leq N.}

By construction, each value of $y_n$ is a (random) estimator of $y(\g{x}_n)=\log\PP{\mathbb{E}_{\g{Z}}\Cr{g(\g{x}_n,\g{Z})}}$, and these $n$ estimators are statistically independent of each other as long as they rely on independent realizations of $\g{Z}$. If we denote by $m(\g{x}_n)$ and $s(\g{x}_n)$ the mean and standard deviation of $g(\g{x}_n,\g{Z})$,

\Eq{m(\g{x}_n):=\mathbb{E}_{\g{Z}}\Cr{g(\g{x}_n,\g{Z})}=\exp\PP{y(\g{x}_n)}, \quad s^2(\g{x}_n):=\text{Var}_{\g{Z}}(g(\g{x}_n,\g{Z})),}

\noindent{}the Central Limit Theorem (CLT) also assures us that 

\Eq{\frac{\sqrt{R}(g_n-m(\g{x}_n))}{s(\g{x}_n)}  \convL   \mathcal{N}(0,1),}

\noindent{}where $\convL$ is the convergence in law, and where for any $(a,b)\in\R\times\R^+$, $\mathcal{N}(a,b)$ denotes the set of Gaussian random variables whose mean and variance coefficients are equal to $a$ and $b$ respectively. In that case, for $R$ sufficiently large, the Delta method \cite{wolter2007taylor} tells us that, for each $1\leq n\leq N$, $y_n$ is close (in distribution) to 

\Eq{y(\g{x}_n)+G_n \frac{s(\g{x}_n)}{m(\g{x}_n)\sqrt{R}},}

\noindent{}with $\g{G}:=(G_1,\ldots,G_N)$ a Gaussian random vector that is independent of $Y$, and whose components are independent and distributed according to the standard Gaussian distribution. In addition, if we write $\g{Y}_N^{\text{noise}} := (Y(\g{x}_n)+G_n {s(\g{x}_n)}/(m(\g{x}_n)\sqrt{R}))_{n=1}^N$, $\g{\mu}:=(\mu(\g{x}_n))_{n=1}^N$, $\g{r}(\g{x}):=(C(\g{x}_n,\g{x}))_{n=1}^N$, $\g{C}:=(C(\g{x}_n,\g{x}_m))_{1\leq n,m\leq N}$ and \\ $\g{\Gamma}=(\delta_{n,m}{s(\g{x}_n)^2} / (m(\g{x}_n)^2 {R}))_{1\leq n,m\leq N}$, with $\delta_{n,m}$ the Kronecker symbol equal to $1$ if $n=m$ and to $0$ otherwise,
we can write:

\Eq{\PP{\begin{array}{c}
Y(\g{x}) \\
\g{Y}_N^{\text{noise}}
\end{array}
}
 \ \sim \ \mathcal{N}\PP{
 \PP{\begin{array}{c}
 \mu(\g{x}) \\
 \g{\mu}
 \end{array}}, 
 \Cr{
\begin{array}{cc}
  C(\g{x},\g{x}) & \g{r}(\g{x})^T \\
  \g{r}(\g{x}) & \g{C} + \g{\Gamma}
\end{array} 
 }
 }.
}

Using the properties of Gaussian conditioning, $Y_N:=\PP{Y \ \vert \ \g{Y}_N^{\text{noise}}=\g{y}}$ is also a Gaussian process, whose mean function $\mu_N$ and covariance function $C_N$ are given by



\Eq{\mu_N(\g{x})=\mu(\g{x}) + \g{r}(\g{x})^T\PP{\g{C}+\g{\Gamma}}^{-1}(\g{y}-\g{\mu}), \ \ \ \g{x}\in \mathbb{X},}

\Eq{C_N(\g{x},\g{x}')= C(\g{x},\g{x}') - \g{r}(\g{x})^T\PP{\g{C}+\g{\Gamma}}^{-1}\g{r}(\g{x}'), \ \ \ \g{x},\g{x}'\in \mathbb{X}.}

Note that even if $Y$ is conditioned by $N$ noisy evaluations of $y$ (due to the MC approximation of the expectation), $Y_N$ approximates the noiseless value of $y(\g{x})$. And under this Gaussian assumption, for each $\g{x}\in\mathbb{X}$, the mean of $Y_{N}(\g{x})$, also called Kriging predictor, constitutes the best approximation of $y(\g{x})=\log(\mathbb{E}_{\g{Z}}\Cr{g(\g{x},\g{Z})})$ in a L$^2$-sense, while its variances $C_N(\g{x},\g{x})$ allows to quantify the approximation error. As $h^\star(\g{x})=c\times \exp(y(\g{x}))$, this also means that $\mathbb{E}\Cr{\exp(Y_{N}(\g{x}))}$ is the best approximation of  $h^\star(\g{x})/c$ in the L$^2$-sense, and that its precision can be characterized by $\text{Var}\Cr{\exp(Y_{N}(\g{x}))}$. Assuming that $\int_{\mathbb{X}} \exp(\widetilde{y}_N(\g{x}'))d\g{x}'<+\infty$ for any realization $\widetilde{y}_N$ of $Y_N$, and defining

\Eq{H_N := \frac{\exp(Y_{N})}{\int_{\mathbb{X}} \exp(Y_{N}(\g{x}'))d\g{x}'},\label{approxHN}}

\noindent{}we can then get rid of the normalization constant $c$, and propose the function 
$\g{x}\mapsto \widehat{h}_N(\g{x}):=\mathbb{E}\Cr{H_N(\g{x})}\label{approxMoyN}$
as an approximation for $h^\star$, whose accuracy in each point of $\mathbb{X}$ can be characterized by:

\Eq{\g{x}\mapsto \widehat{V}_N(\g{x}):=\text{Var}\PP{H_N(\g{x})}.\label{approxVarN}}

The suitability of $Y_N$ to approximate the function $y$ relies heavily on the choice of the statistical moments of $Y$. As the meticulous optimization of these functions is not at the heart of this work, we propose here to set them to standard choices of the literature. Hence, the function $\mu$ is chosen as a constant, the function $C$ is taken in the class of tensorized {Mat\'ern} kernels with smoothing parameter $\nu=5/2$ (the Mat\'ern 5/2 kernel is selected here because it performs well and it is common in the literature \cite{leriche2021revisiting}, but alternative classes of functions can be found in \cite{Santner2003}):


\Eq{\mu(\g{x}):=\beta, \ C(\g{x},\g{x}'):=\sigma^2  \prod_{i=1}^{d} \left(1+\sqrt{5}u_i+\frac{5}{3}u_i^2\right)\exp\left(-\sqrt{5}u_i\right), \ u_i:=\frac{\vert x_i-x_i' \vert }{\ell_i}, \label{momentsY}}


\noindent{}and the hyperparameter vector $\g{\theta}:=(\beta,\sigma,\ell_1,\ldots,\ell_d)$ is estimated by its maximum likelihood estimator (see \cite{Williams2006} for more details).


$\bullet$ \textbf{Coupling GPR and MCMC.} Since the function $\widehat{h}_N$ is defined in the form of an expectation, a first approach to its construction is to generate a set of $M\gg 1$ independent realizations of $Y_N$, denoted $\widetilde{y}_N^{(1)},\ldots,\widetilde{y}_N^{(M)}$, compose these realizations with the exponential function, calculate their integrals over $\mathbb{X}$, which we write

\Eq{\widetilde{c}_N^{(m)}:= \int_{\mathbb{X}} \exp(\widetilde{y}_N^{(m)}(\g{x}'))d\g{x}', \ \ 1\leq m\leq M,\label{defIntegCn}}

\noindent{}and finally take the empirical mean of their ratio:

\Eq{\widehat{h}_{N,M}(\g{x}):= \frac{1}{M}\sum_{m=1}^{M} \frac{\exp\PP{\widetilde{y}_N^{(m)}(\g{x})}}{\widetilde{c}_N^{(m)}}, \quad \g{x}\in\mathbb{X}. }

Note, however, that the complexity associated with projecting a realization of $Y_N$ onto $Q$ points is in $Q^3$ (using a Cholesky algorithm for modeling spatial correlation, for example). It is therefore not possible to project these realizations of $Y_N$ onto a too large number of points in $\mathbb{X}$, which makes it very difficult (or even impossible when the dimension of the input space increases) to try and approximate these integrals by discretizing $\mathbb{X}$. On the contrary, it seems more reasonable to try to project $Y_N$ only in the zones where it is likely to take large values, i.e. the zones that are likely to make a non-negligible contribution to the integrals presented in Eq. (\ref{defIntegCn}). And to do this, we propose to couple this generation of realizations of $Y_N$ to MCMC-type sampling techniques. 
In order to limit the number of projection points and maximize the acceptance rate of a new point in the MCMC procedure, we propose to turn to algorithms that exploit the derivative of the function of interest. 
Indeed,
once Gaussian predictor $Y_N$ has been calculated, it is interesting to note that we also have access to its gradient $\nabla Y_N$ by simply deriving the obtained mean and covariance functions:

\Eq{\nabla  Y_N(\g{x}) \ \sim \ \text{GP}\PP{\nabla  \mu_N,\nabla^2C_N},}

\Eq{(\nabla  Y_N(\g{x}))_i := \frac{\partial Y_N}{\partial x_i}(\g{x}), \quad (\nabla  \mu_N(\g{x}))_i := \frac{\partial \mu_N}{\partial x_i}(\g{x}), \ \ 1\leq i\leq d,}

\Eq{ 
(\nabla^2C_N(\g{x},\g{x}'))_{i,j}:=\frac{\partial^2 C_N}{\partial x_i \partial x_j}(\g{x},\g{x}'), \ \ 1\leq i,j\leq d.
}

Note that choosing a Mat\'ern 5/2 kernel as the covariance function for $Y$, which is precisely four times differentiable, assures us that the derivatives of $C_N$ are well defined.
In order to take full advantage of this gradient, we then propose to turn to the Metropolis-adjusted Langevin algorithm (MALA) (see Algorithm SM1 
in Supplementary Material for a synthetic description of this algorithm) \cite{roberts1996exponential,roberts1998optimal}. This method, which can be seen as an Hamiltonian Monte Carlo algorithm with only one discrete time step, couples two mechanisms: a Langevin dynamics that drives the random walk towards regions of high probability, and a Metropolis–Hastings accept/reject procedure that improves the mixing and convergence properties of the random walk. 
Using the same notations as in the description of the MALA algorithm in Supplementary Material, Algorithm \ref{algoMALA_GPR} is therefore proposed to couple the MALA algorithm, initially designed to work with deterministic functions, with the sequential construction of a particular trajectory of $Y_N$.

\begin{algorithm}[H]
\caption{Application of the MALA algorithm to a realization of $Y_N$.}
\begin{algorithmic}[1]
\STATE{Let $Z \sim \text{GP}(\mu_N,C_N)$ be a copy of $Y_N$, $\nabla Z$ be its gradient, $\g{a}_1$ be an element \\ of $\mathbb{X}$, $\tau>0$ be a chosen constant, and $K$ be a maximum number of steps. }
\STATE{Let $(z_1,\nabla z_1)\in\R^{d+1}$ be a particular realization of $(Z(\g{a}_1),\nabla Z(\g{a}_1))$.}
\STATE{Update: $Z \ \leftarrow Z \ \vert \ (Z(\g{a}_1),\nabla Z(\g{a}_1))=(z_1,\nabla z_1)$. }
\STATE{For all $\g{x},\g{x}'$ in $\mathbb{X}$ and all $\nabla {z}$ in $\R^{d}$, define $q(\g{x}' \vert \g{x},\nabla z):= -{\frac {1}{4\tau }}\NOR{\g{x}'-\g{x}-\tau \nabla z }^{2}$.}
\FOR{$k \in \Ac{1,\ldots,K-1}$}
\STATE{Compute $\widetilde{\g{a}}:= \g{a}_{k} + \tau \nabla z_k + \sqrt{2\tau} \g{\xi}_{k}$, with $\g{\xi}_k \sim\mathcal{N}(\g{0},\g{I}_{d})$. }

\STATE{Let $(\widetilde{z},\nabla \widetilde{z})\in\R^{d+1}$ be a particular realization of $(Z(\widetilde{\g{a}}),\nabla Z(\widetilde{\g{a}}))$.}
\STATE{Update: $Z \ \leftarrow Z \ \vert \ (Z(\widetilde{\g{a}}),\nabla Z(\widetilde{\g{a}}))=(\widetilde{z},\nabla \widetilde{z})$. }
 \STATE{Compute $  \alpha := \min \PP{ 0, \widetilde{z}- z_k + q(\g{a}_{k} \vert \widetilde{\g{a}},\nabla \widetilde{z}) - q( \widetilde{\g{a}} \vert \g{a}_{k},\nabla z_k)} .$
 }
\STATE{Sample at random $U_k$ uniformly in $[0,1]$.
 }
\IF{$\alpha \geq \log(U_k)$} 
\STATE{(Accept) Set  $\g{a}_{k+1}:=\widetilde{\g{a}} $, $z_{k+1}:=\widetilde{z}$, $\nabla z_{k+1}:= \nabla \widetilde{z}$.}
\ELSE 
\STATE{(Reject) Set  $\g{a}_{k+1}:=\g{a}_k $.}
\ENDIF
\ENDFOR
\RETURN $\g{a}_1,\ldots, \g{a}_K$.
\end{algorithmic}
\label{algoMALA_GPR}
\end{algorithm} 

If we note $\gamma$ the average rate of acceptance of the new points within Algorithm \ref{algoMALA_GPR}, it is important to note that launching this algorithm with a maximum of $K$ steps requires being able to condition the process $Y_N$ by on average $(d+1)\gamma K$ values (remember that the dimension of $\mathbb{X}$ is noted $d$), which can quickly become too costly numerically. Indeed, for each accepted point, the process is supposed to be conditioned by a value of $y$, but also by $d$ values of its derivatives. Several heuristics can therefore be proposed to reduce the numerical cost of this algorithm without greatly altering its results. For example, we can propose to condition $Y_N$ only by the values of $y$, and not by the values of its derivatives, noting that the points returned in the Markov chain are likely to be two by two relatively close to each other, and that this information on the derivatives will be implicitly encoded in the sequences of $y$ values. Alternatively, the total number of conditioning points can be limited to $Q$. For this, at each increment $k$ of Algorithm \ref{algoMALA_GPR}, from the moment the number of accepted points exceeds $Q$, we can propose to replace the process $Z$ in line 8 of Algorithm \ref{algoMALA_GPR} by the process 

\Eq{ Y_N \ \vert \  (Y_N(\g{a}_{j_1}),\nabla Y_N(\g{a}_{j_1}))=(\widetilde{z}_{j_1},\nabla \widetilde{z}_{j_1}),\ldots, (Y_N(\g{a}_{j_Q}),\nabla Y_N(\g{a}_{j_Q}))=(\widetilde{z}_{j_Q},\nabla \widetilde{z}_{j_Q}),} 

\noindent{}where for each $1\leq q\leq Q$, the values of $\widetilde{z}_{j_q}$ and $\nabla \widetilde{z}_{j_q}$ correspond to the values taken by the processes $Z$ and $\nabla Z$ in $\g{a}_{j_q}$ at iteration $j_q$. In order to choose the points $\g{a}_{j_1},\ldots,\g{a}_{j_Q}$, we can note that the larger $C_N(\g{x},\g{x}')$ is, the more the weight of the observations in $\g{x}'$ will be important in predicting $Y_N$ at $\g{x}$. 
At iteration $k$, this encourages us to take for these points the $Q$ distinct values of $\g{a}_1,\ldots,\g{a}_{k}$ associated with the highest values of $C_N(a_k,a_1),\ldots,C_N(a_k,a_{k})$ (which can be seen as a selection of the $Q$ closest neighbors of $\g{a}_k$ seen by the covariance filter). In this work, we opt for a third possibility, which is based on the principle that conditioning the Gaussian process with a new point only really makes sense if it \textit{really} modifies its statistical properties. To this end, we propose to add a new point $\g{x}$ to the conditioning set only if 

\Eq{ \NOR{(z,\nabla z)-\mathbb{E}\Cr{(Z(\g{x}),\nabla Z(\g{x}))}} }

\noindent{}is bigger than a prescribed threshold $\gamma>0$, where $(z,\nabla z)$ is the sampled value of $(Z(\g{x}),\nabla Z(\g{x}))$ at line 7 of Algorithm \ref{algoMALA_GPR}.

$\bullet$ \textbf{Non-parametric reconstruction of the trajectories of $H_N$.}
Running the former algorithm $M$ times independently produces $M$ sequences of $K$ points. These points can be post-processed (removal of the burn-in phase, sub-sampling to limit potential correlation, see \cite{Rubinstein2008} for further details about the analysis of MCMC results) in order to extract a set of $\widetilde{K}\ll K$ points that can be considered statistically independent and approximately sampled according to the PDFs associated with the trajectories of $H_N$. Then, if we restrict ourselves to non-parametric reconstruction methods such as kernel density estimation (KDE) \cite{perrinCSDA2017}, and if we denote these points by $\{\widetilde{\g{x}}_{1}^{(1)},\ldots,\widetilde{\g{x}}_{\widetilde{K}}^{(1)}\}$,$\ldots$,$\{\widetilde{\g{x}}_{1}^{(M)},\ldots,\widetilde{\g{x}}_{\widetilde{K}}^{(M)}\}$, we can approximate the trajectories of $H_N$ in the following form:

\Eq{\widehat{f}^{(m)}_{N,\widetilde{K}}(\g{x}) := \frac{1}{\widetilde{K} \text{det}(\g{B}_m) } \sum_{k=1}^{\widetilde{K}}  \kappa_{m}\PP{\g{B}_m^{-1}(\g{x}-\widetilde{\g{x}}_{k}^{(1)})}, \ \ 1\leq m\leq M, \label{approxPDF_np}}

\noindent{}with $\g{B}_m$ an invertible matrix and $\kappa_m$ a positive function whose integral over $\mathbb{X}$ is $1$. The choice of $\g{B}_m$ is particularly central for this kind of approximation. Indeed, too large values tend to smooth out and erase the singularities in the reconstruction of the trajectory of $H_N$, while too small values can artificially increase the number of modes in it. Many works can be found in the literature that focus on the best estimate of $\g{B}_m$. In this work, we limit ourselves to the case where
$\g{B}_m = \omega_m \g{R}^{1/2}_m
$, with $\omega_m$ estimated using an adapted likelihood maximization procedure \cite{perrinCSDA2017} and

\Eq{\g{R}_m^{1/2}(\g{R}_m^{1/2})^T := \frac{1}{\widetilde{K}-1} \sum_{k=1}^{\widetilde{K}} \PP{\widetilde{\g{x}}_{k}^{(m)}-  \sum_{\ell=1}^{\widetilde{K}} \frac{\widetilde{\g{x}}_{\ell}^{(m)}}{\widetilde{K}-1} }\PP{\widetilde{\g{x}}_{k}^{(m)}-  \sum_{\ell=1}^{\widetilde{K}} \frac{\widetilde{\g{x}}_{\ell}^{(m)}}{\widetilde{K}-1} }^T.}

The other parameter to be adjusted in this approximation is the kernel function. Many choices are again possible \cite{tsybakov2004introduction}, the most frequent choice when $\mathbb{X}=\R^{d}$ being the set of Gaussian kernels, so that $\kappa_m$ is the PDF of a Gaussian standard variable.
These $M$ non-parametric reconstructions finally lead us to the definition of the following estimator for $h^\star(\g{x})$ for each $\g{x}\in\mathbb{X}$ :

\Eq{
\widetilde{h}_{N,M,\widetilde{K}}(\g{x})  := \frac{1}{M}\sum_{m=1}^M \widehat{f}_{N,\widetilde{K}}^{(m)}(\g{x}) 
 = \frac{1}{M\widetilde{K}}\sum_{m=1}^M \frac{1}{ \text{det}(\g{B}_m) } \sum_{k=1}^{\widetilde{K}}  \kappa_{m}\PP{\g{B}_m^{-1}(\g{x}-\widetilde{\g{x}}_{k}^{(m)})}.
\label{estimatorFinal}}

$\bullet$ \textbf{Sequential uncertainty reduction.}
Just as $\widetilde{h}_{N,M,\widetilde{K}}(\g{x})$ approaches $\widehat{h}_N(\g{x}) = \mathbb{E}\Cr{H_N(\g{x})}$, the precision of such a reconstruction of $h^\star$ can thus be characterized by 

\Eq{\widetilde{V}_{N,M,\widetilde{K}}(\g{x}) := \frac{1}{M-1}\sum_{m=1}^M \PP{ \widehat{f}^{(m)}_{N,\widetilde{K}}(\g{x}) - \widetilde{h}_{N,M,\widetilde{K}}(\g{x}) }^2, \label{defVarh_tilde}}

\noindent{}as an approximation of $\widehat{V}_N(\g{x})=\text{Var}(H_N(\g{x}))$, where we recall that $H_N$ has been defined in Eq. (\ref{approxHN}). 
We note the presence in this variance of the three constants $\widetilde{K}$, $N$ and $M$, whose meanings are very different. By increasing $\widetilde{K}$ (which amounts to increasing the maximum number of steps $K$ of Algorithm \ref{algoMALA_GPR}), we hope to estimate each trajectory of $H_N$ better and better. Note however that despite the presence of $\widetilde{K}$, the reconstruction error of each trajectory of $H_N$ by its KDE is not truly integrated into this variance. In order to empirically characterize this variance related to $\widetilde{K}$, one could for example use Bootstrap techniques on the $\widetilde{K}$ available values of $\g{x}$ during the KDE of each trajectory of $H_N$, the value of $\widetilde{K}$ being finally chosen to ensure a sufficiently low reconstruction variance. Then, by increasing $M$, it is the expectation and the variance of $H_N$ that we seek to make converge to their true values. Here again, it is possible to adapt the value of M to ensure that the variance of the empirical estimator of the expectation, i.e. $\widetilde{V}_{N,M,\widetilde{K}}(\g{x})/M$, is sufficiently small.
At last, by increasing $N$, it is the approximation of $y$ that we try to make more precise. And this time it would be the convergence of the sequence $(\widetilde{V}_{N,M,\widetilde{K}}(\g{x}))_{N\geq 1}$ that we could study to know when to stop the enrichment. For this work, as we are working with a limited total budget, it is ultimately not really when to stop that will interest us, but rather how to choose where to add the new code evaluations so that this variance is as low as possible once the total budget is reached. This question of the localization of the new evaluation of the code in a GPR framework is at the heart of many works in the literature, whether for optimization \cite{jones2001taxonomy,frazier2018tutorial,gramacy2020surrogates}, reliability analysis \cite{Bichon2011,Fauriat,Bect2012,Chevalier2014,PerrinRESS2016} or even code calibration \cite{Damblin2018,perrinRESS2019}. By applying the results of these works to our case, we can then add new points sequentially, by choosing as the new value of $\g{x}$ to explore the value that maximizes $\widetilde{V}_{N,M,\widetilde{K}}$, 

\Eq{\g{x}_{N+1} := \arg\max_{\g{x}\in\mathbb{X}} \widetilde{V}_{N,M,\widetilde{K}}(\g{x}).}

Note that this selection criterion seeks to minimize the absolute errors on $h^\star$ rather than the relative errors, in the sense that we give little importance to the areas of low value for $h^\star$. This is consistent with the type of application targeted for this work, namely estimating \textit{a priori} distributions in a Bayesian framework, but other selection criteria could of course be proposed for other types of application. Solving the optimization problem on the entire domain $\mathbb{X}$ can nevertheless be complicated in practice. We can thus propose a simplified version of this optimization problem by limiting the search to the set $\{ \widetilde{\g{x}}_k^{m}, \ 1\leq m\leq M, \ 1\leq k\leq \widetilde{K} \}$ that gathers the points provided by the $M$ runs of the MALA algorithm on the $M$ different trajectories of $H_N$.

\subsection{Summary of the proposed method and general remarks}
\label{sec23}

In summary, the approach we propose to solve the reconstruction problem presented in Eq. (\ref{pb_mini}) is based on the coupling of three types of methods: (1) MC sampling methods for the approximation of mathematical expectations and MCMC sampling methods for the generation of points from a target PDF, (2) Gaussian process regression (with heteroscedastic noise taken into account) for the approximation of the logarithm of the function to be reconstructed, and (3) a non-parametric kernel reconstruction approach for the reconstruction of PDFs from samples drawn from these PDFs. The proposed method not only provides an approximation of the function of interest, but also characterizes its accuracy in the form of a reconstruction variance. This variance is mainly related to the number $N$ of points used for the GPR approximation, the number $\widetilde{K}$ of MCMC draws that can be considered statistically independent, and the number $M$ of GPR trajectories considered. By increasing the computational cost, i.e. by increasing $N$, $\widetilde{K}$ and $M$, it is possible to reduce this variance, in a more or less efficient way (e.g. by optimizing the selection of new points computed for the enrichment of the GPR model). This whole process is finally summarized in Algorithm SM2 
in Supplementary Material.

\section{Application}
\label{sec3}

The objective of this section is to emphasize the possibilities brought by the proposed method to efficiently solve reconstruction problems that can be written under the form of Eq. (\ref{pb_mini}). To this end, we present two applications. The first one is based on an analytic model in dimension $d=2$, for which function $h^\star$ is known, which allows us to illustrate most of the results graphically, and to highlight most of the difficulties associated with reconstructing $h^\star$ in a completely controlled context. The second application concerns the identification in a Bayesian framework of the mechanical and energy properties of a high-speed train from a set of acceleration and power measurements. The main difficulty in this reconstruction arises from the fact that the driver commands are only partially known on the different measured train runs. In that case, considering the true driver commands as random, the inference problem can be written in the form of Eq. (\ref{pb_mini}), as it is explained in Section \ref{sec21}.

\subsection{Application on an analytical example}

The proposed approach is first illustrated
on a numerical application based on simulated data. In order to be as consistent as possible with the theoretical framework introduced in Section \ref{sec21}, we seek to construct an estimator of $h^\star$ such that for all $\g{x}\in \mathbb{X}:=[-1,8]^2$,

\Eq{h^{\star}(\g{x})=\frac{\mathbb{E}_{{Z}}\Cr{g(\g{x},Z)}}{\int_{\mathbb{X}} \mathbb{E}_{{Z}}\Cr{g(\g{x}',Z)} d\g{x}'},}

\noindent{}where $Z\sim\mathcal{N}(0,1)$ is a standard Gaussian variable, and where for each $z$ in $\R$,

\Eq{
g(\g{x},z) = \exp\PP{
-\frac{x_1^2x_2^2+x_1^2+0.95x_2^2-8x_1-8x_2}{2}
}\times \exp\PP{z-\frac{1}{2}}
.
}

\BFig{
\subfigure[ ]{
\psfrag{x}[c][c][1]{$x_1$}
\psfrag{y}[c][c][1]{$x_2$}
\includegraphics[height = 5.5cm,keepaspectratio]{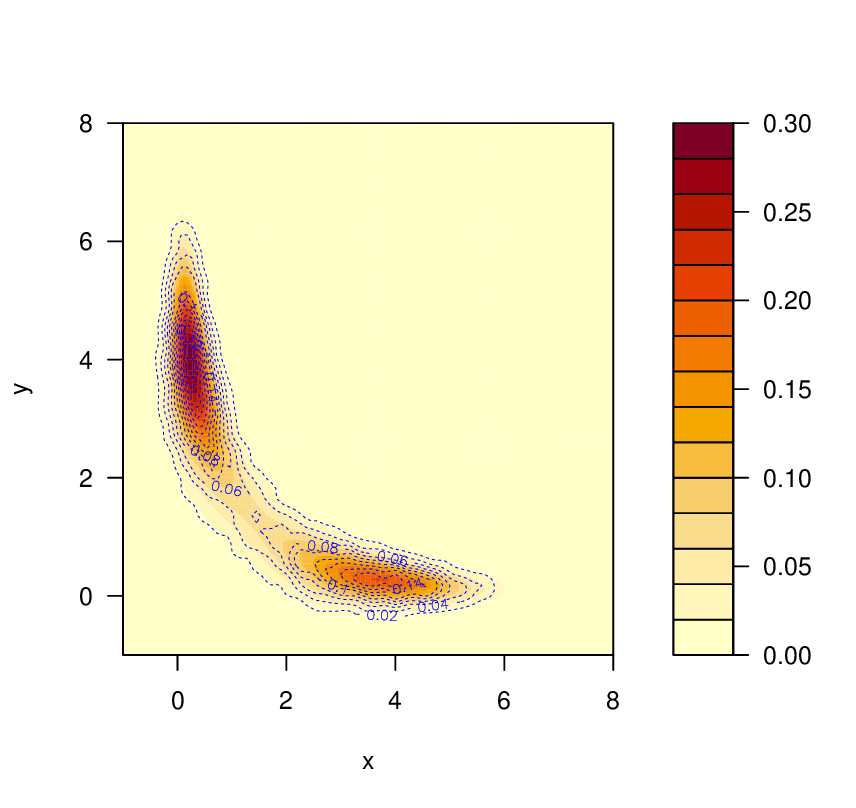}}
\subfigure[ ]{
\psfrag{x}[c][c][1]{$x_1$}
\psfrag{y}[c][c][1]{$\int_{\R}h^\star(x_1,x_2)dx_2$}
\includegraphics[height = 5.5cm,keepaspectratio]{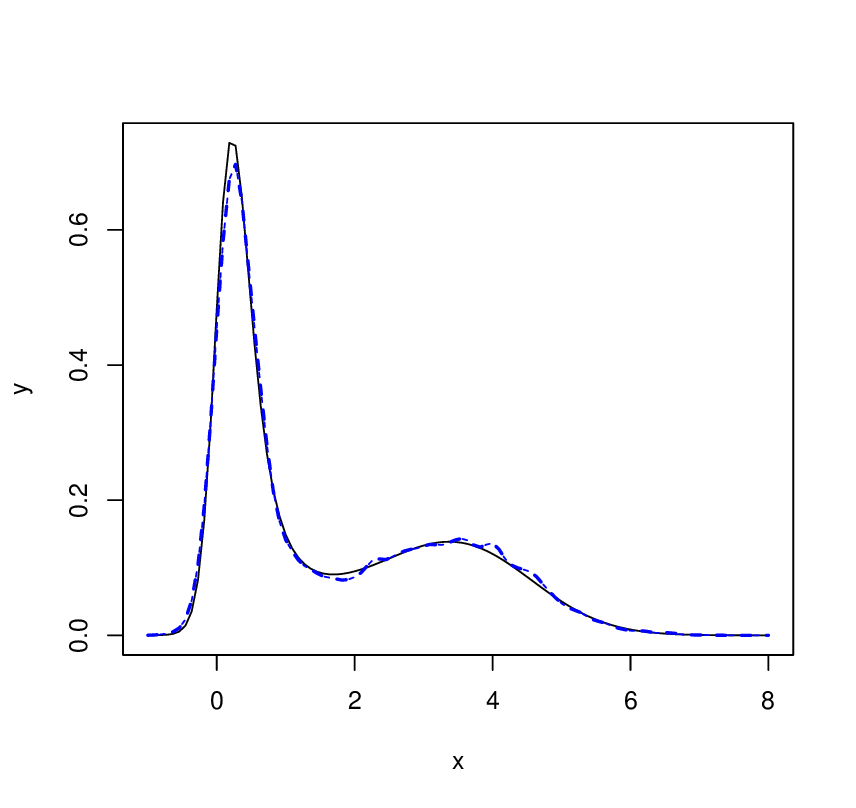}}
\subfigure[ ]{
\psfrag{x}[c][c][1]{$x_2$}
\psfrag{y}[c][c][1]{$\int_{\R}h^\star(x_1,x_2)dx_1$}
\includegraphics[height = 5.5cm,keepaspectratio]{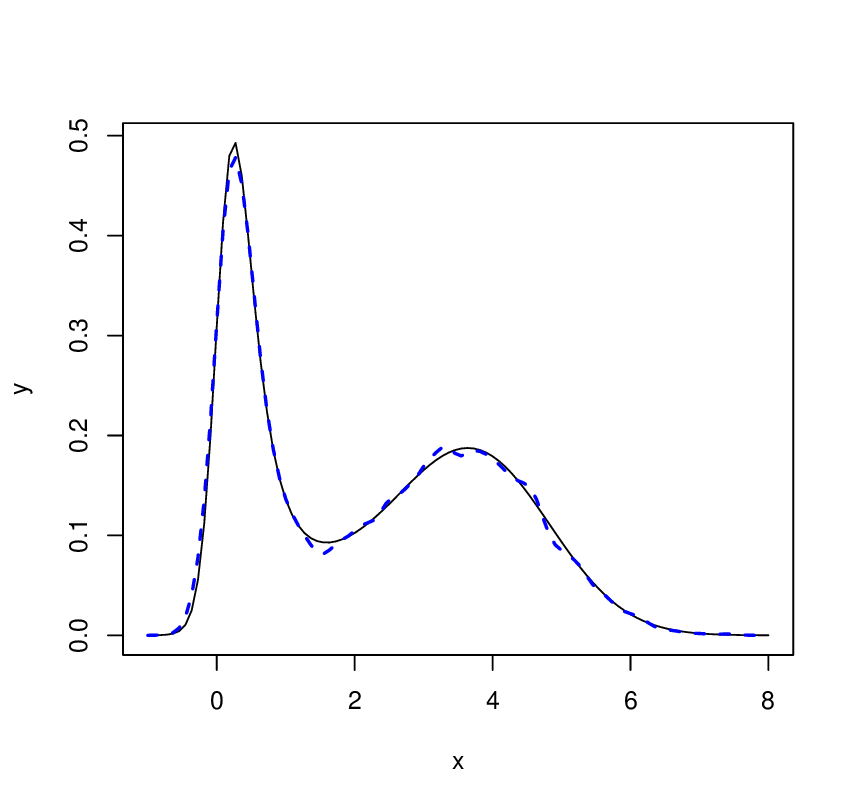}}
\caption{Graphical representations of function $\g{x}\mapsto h^{\star}(\g{x})$ and of its approximation when coupling the MALA algorithm with a KDE reconstruction relying on $\widetilde{K}=10^4$ points. The true function $h^{\star}$ corresponds to the filled contour lines in subfigure (a) and to the black solid lines in subfigures (b) and (c), while its approximated version correspond to the blue dotted lines.}
\label{fig1_fonction}
}

Note that function $h^{\star}$, whose contour plot is shown in Figure \ref{fig1_fonction}, deliberately has two main modes (to complicate its reconstruction), and that its values are close to zero over a large part of $\mathbb{X}$, which is a configuration very often encountered in Bayesian inversion. Note also that the constant $\int_{\mathbb{X}} \mathbb{E}_{{Z}}\Cr{g(\g{x}',Z)} d\g{x}'$ is assumed unknown (in the sense that it is too costly to estimate) in the following. For this analytical example, it is clear that the calculation of $\mathbb{E}_{{Z}}\Cr{g(\g{x},Z)}$ is explicit (and therefore very inexpensive). This makes it possible to check, as a first step, the efficiency of the (MALA+KDE) pair for reconstructing $h^\star$ without needing to calculate the normalization constant. The MALA algorithm presented implemented with function $\log(\mathbb{E}_{{Z}}\Cr{g(\g{x},Z)})$ first generates points which can be considered as i.i.d. and which are supposed to come from $h^\star$ when seen as a PDF. Then the reconstruction using KDE approches provides, from these generated points, an estimate of $h^\star$ at any point of input space $\mathbb{X}$. As indicated in Section \ref{sec22}, this reconstruction is highly dependent on the number of i.i.d. points generated, which we noted $\widetilde{K}$. In the following, for the sake of consistency of results, all KDE constructions will be based on $\widetilde{K}=10^4$ points. The relevance of such a reconstruction for this value of $\widetilde{K}$ can also be seen in Figure \ref{fig1_fonction}, in terms of both joint and marginal distributions. 
If we now assume that each evaluation of $g$ is numerically costly, it is no longer possible to proceed as previously. In order to enable a much more cost-effective reconstruction of $h^\star$, we now propose to follow the approach proposed in Section \ref{sec2}.  
Using $N_0\times R$ i.i.d. realizations of $Z$, which are noted ${z}_{1,1},\ldots,{z}_{1,R},{z}_{2,1}\ldots,{z}_{N_0,R}$, we therefore denote by $g_1,\ldots,g_{N_0}$ the $N_0$ independent estimators of $\mathbb{E}_{{Z}}\Cr{g(\g{x}_1,Z)},\ldots,\mathbb{E}_{{Z}}\Cr{g(\g{x}_{N_0},Z)}$ in the $N_0$ particular values $\g{x}_1,\ldots,\g{x}_{N_0}$ of $\g{x}$ in $\mathbb{X}$, so that:

\Eq{g_n=\frac{1}{R}\sum_{r=1}^R g(\g{x}_n,z_{n,r}), \ \ \ 1\leq n\leq N_0. }

For all $1\leq n\leq N_0$, the associated estimators of $y(\g{x}_n):=\log\PP{\mathbb{E}_{{Z}}\Cr{g(\g{x}_n,Z)}}$ are then noted $y_n:=\log(g_n)$. The approximate Gaussian behavior of these estimators when $R$ is chosen sufficiently high is illustrated in Figure \ref{compGaussFig}, which compares (here for $R=100$) the histograms of the values of $y_1$ and $y_2$ that we would  obtain when independently repeating their estimation a high number of times with the PDFs of Gaussian random variables with means $y(\g{x}_1)$ and $y(\g{x}_2)$ and same variance $$\frac{1}{R}\frac{s^2(\g{x}_n)}{m^2(\g{x}_n)}=\frac{1}{R}\frac{\text{Var}_Z\PP{g(\g{x}_n,{Z})}}{\mathbb{E}_Z\Cr{g(\g{x}_n,{Z})}^2}=\frac{\exp(1)-1}{R}.$$




The very good match observed between these empirical dispersions (the same match is observed for the other values of $\g{x}$) and the associated Gaussian distributions leads us to model the statistical noise on the estimates of $y(\g{x}_1),\ldots,y(\g{x}_n)$ in the form of a centred Gaussian noise with a variance independent of $\g{x}$ equal to $(\exp(1)-1)/R$.

\BFig{
\psfrag{titre}[c][c][1]{}
\psfrag{x}[c][c][1]{Values of $y_1$}
\psfrag{y}[c][c][1]{PDF}
\subfigure[$\g{x}_1=(1,2)$]{
\includegraphics[scale=0.4]{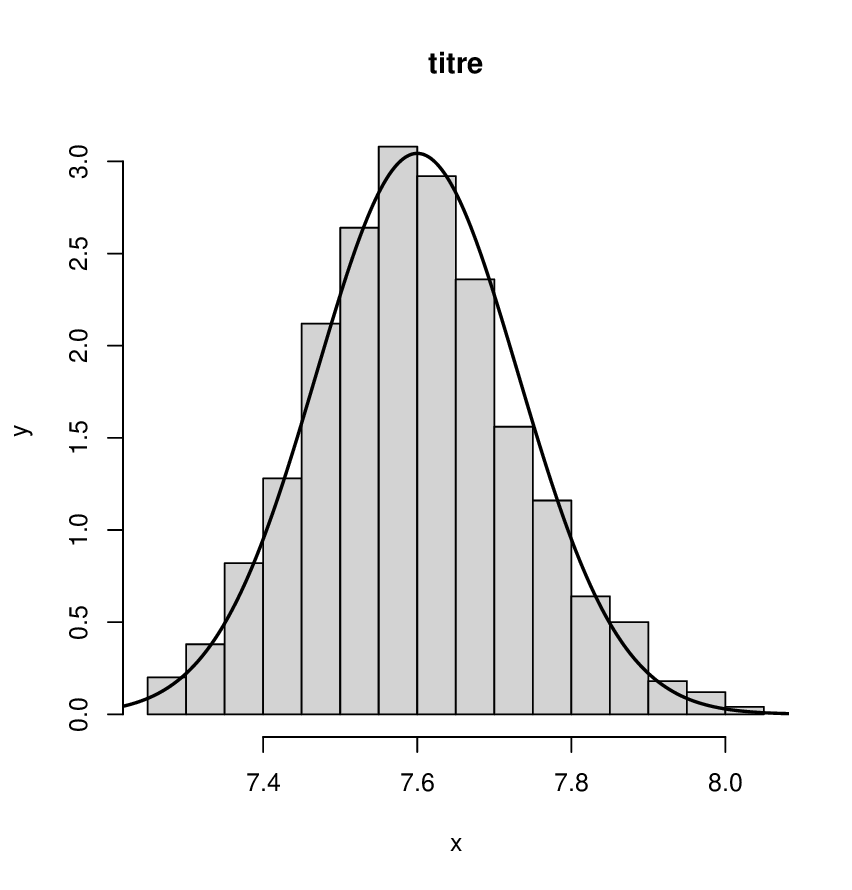}
}
\subfigure[$\g{x}_2=(3,6)$]{
\psfrag{titre}[c][c][1]{}
\psfrag{x}[c][c][1]{Values of $y_2$}
\psfrag{y}[c][c][1]{PDF}
\includegraphics[scale=0.4]{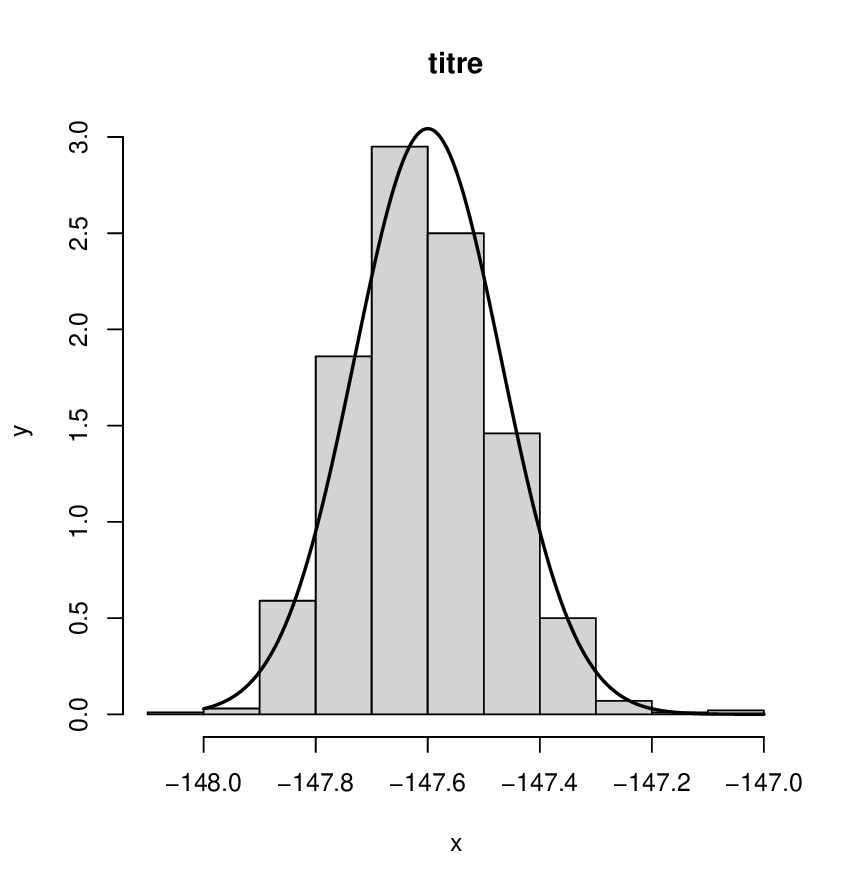}
}
\caption{Graphical assessment of the Gaussian behavior of the empirical estimator of $\log(\mathbb{E}_{{Z}}\Cr{g(\g{x},Z)})$ using $R$ independent realizations of $Z$ for two particular values of $\g{x}$ ($\g{x}_1=(1,2)$ and $\g{x}_2=(3,6)$). In both figures, the histograms characterize the observed fluctuations of the values of $y_1$ when repeating the generations of $Z$, while the black solid lines correspond to Gaussian PDF with means $\log(\mathbb{E}_{{Z}}\Cr{g(\g{x}_1,Z)})$ and $\log(\mathbb{E}_{{Z}}\Cr{g(\g{x}_2,Z)})$ and same variances $(\exp(1)-1)/R$.}
\label{compGaussFig}
}

Given these $N_0$ noisy observations of $y:\g{x}\mapsto y(\g{x})=\log\PP{\mathbb{E}_{{Z}}\Cr{g(\g{x},{Z})}}$, and following the formalism presented in Section \ref{sec22}, function $y$ is now modeled by the conditioned Gaussian process $Y_{N_0}$, whose mean function $\mu_{N_0}$ and covariance function $C_{N_0}$ are fully parametrized (see Eq. (\ref{momentsY})) by a vector $(\beta,\sigma^2,\ell_1,\ell_2)$, which we choose to estimate by maximum likelihood. As explained in Section \ref{sec22}, we can then couple this Gaussian approximation of $y$ to a MALA-type MCMC algorithm to approximate $h^\star$. By choosing $R=100$, and using only $N_0=20$ noisy estimates of $y$ (which corresponds to $20 \times R = 2000$ evaluations of the code $g$), Figure \ref{figCompMoyGPR} compares, for example, the contourplots of $y$ and $\mu_{N_0}$, as well as the associated contours of $h^\star$ and $$\widehat{h}_{\mu_{N_0}}:=\frac{\exp(\mu_{N_0})}{\int_{\mathbb{X}}\exp(\mu_{N_0}(\g{x}'))d\g{x}'}.$$ 
It should be noted that function $\widehat{h}_{\mu_{N_0}}$ is constructed by kernel density estimation (KDE) relying on $\widetilde{K}=10,000$ independent points that can be considered statistically independent and approximately sampled according to $\widehat{h}_{\mu_{N_0}}$ when running the MALA algorithm with function $\widehat{h}_{\mu_{N_0}}$ instead of $h$ (with $K \gg \widetilde{K}$). In this way, we can be sure that $\int_{\mathbb{X}}\widehat{h}_{\mu_{N_0}}(\g{x}')d\g{x}'=1$ without having to calculate $\int_{\mathbb{X}}\exp(\mu_{N_0}(\g{x}'))d\g{x}'$. Nevertheless, for this analytical case, we know the true form of $h^\star$, and we can see that the approximation obtained is relatively poor: the large values of $\widehat{h}_{\mu_{N_0}}$ are concentrated on one (or two) modes that do not coincide with the true modes of $h^\star$. We could also have reached the same conclusion by evaluating the accuracy of such a construction by not running one MCMC procedure on the mean of $Y_{N_0}$, but on a series of $M$ independent trajectories of $Y_{N_0}$, as explained in Algorithm \ref{algoMALA_GPR}. In this case, we obtain $M$ independent approximations of $h^\star$, whose empirical means and variances (or standard deviations) can be calculated at any point of $\mathbb{X}$. However, on the basis of only $N_0$ noisy evaluations of $Y_{N_0}$, it turns out that these estimates of $h^\star$ vary greatly from one trajectory of $Y_{N_0}$ to another, as shown in Figure \ref{figCompMoyGPR}-(c), which plots the standard deviation of these trajectories at any point on $\mathbb{X}$. 
The reason for this very uncertain approximation is clearly linked to the lack of observation points for $g$ and $y$. Two possibilities for adding these new points are compared below. Firstly, we propose to fill the input space as fully as possible, and we call this enrichment procedure the \emph{space-filling} approach. Secondly, we propose adding the new points where the prediction variance is maximal, and we call this second enrichment procedure \emph{var-based}. Figure \ref{figCompStrategieAdd} then compares the reconstructions obtained in the two cases, that is to say the evolutions of $\widetilde{h}_{N,M,\widetilde{K}}$, which is defined by Eq. (\ref{estimatorFinal}), as a function of the number of points added according to these two strategies. 
We then observe, unsurprisingly, that the \emph{Var-based} approach better concentrates the new points where the values of $h^\star$ are most likely to be high without being too close to the areas already assessed, for a reconstruction of $h^\star$ as accurate as possible. For a more quantitative analysis, Figure \ref{boxplot_analy}-(a) compares the evolution of the reconstruction errors, defined as the Wasserstein distances of order 1 between $h^\star$ and $\widetilde{h}_{N,M,\widetilde{K}}$. The procedure being repeated $5$ times with $5$ initial designs of experiments of size $20$, these error evolutions are then presented in the form of boxplots in order to characterize the potential dispersion of results. 
This figure allows us, first, to check that the more information we add about $y$, the more the reconstruction error decreases, until it tends to the reference reconstruction error associated with the approximation of Figure \ref{fig1_fonction}, which was obtained by coupling a KDE approximation of $h^\star$ relying on 10,000 independent points stemming from MCMC procedures on the true (and noiseless) function $y$. Thus, for both sampling strategies, it seems possible to approximate $h^\star$ from a relatively reasonable number of calls to the costly code $g$. 
The advantage of concentrating new points in places where the estimation variance is highest is also clearly visible in this figure, where the error associated with the \emph{var-based} approach decreases much more rapidly than that associated with the \emph{space-filling} approach. Focusing on the \emph{var-based} approach, we can also see that from $N\geq 70$, the reconstruction error is close to the reference error, this tendency being relatively independent of the initial experimental design, as convergence is almost identical for the five repetitions. 

\BFig{
\subfigure[Approximation of $\g{x}\mapsto y(\g{x})$]{
\psfrag{x}[c][c][1]{$x_1$}
\psfrag{y}[c][c][1]{$x_2$}
\includegraphics[height=5.5cm]{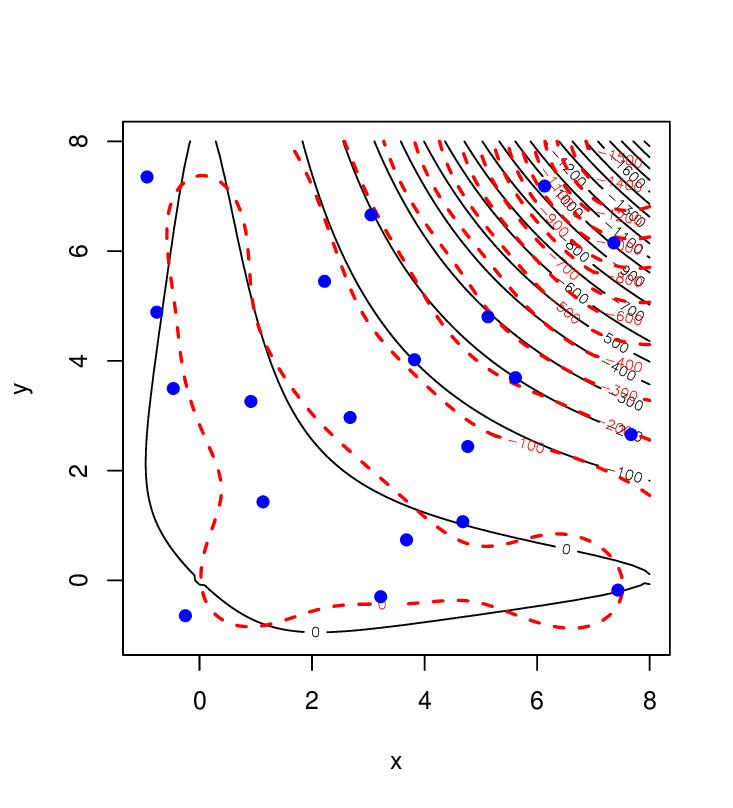}}
\subfigure[Associated estimation of $h^\star=\frac{\exp(y)}{\int_{\mathbb{X}}\exp(y(\g{x}'))d\g{x}' }$]{
\psfrag{x}[c][c][1]{$x_1$}
\psfrag{y}[c][c][1]{$x_2$}
\includegraphics[height=5.5cm]{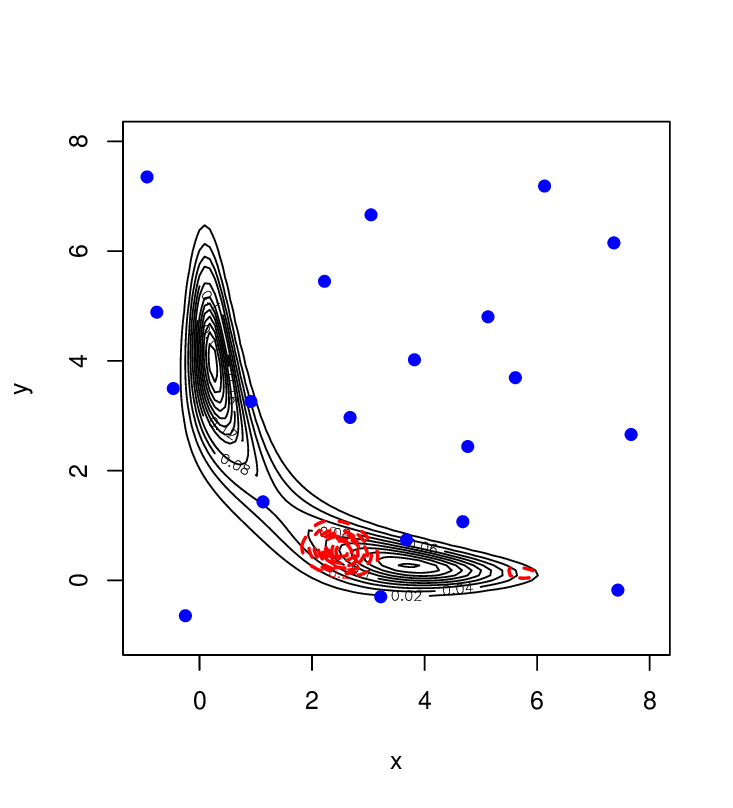}}
\subfigure[Standard deviation of the estimate of $h^\star$]{
\psfrag{x}[c][c][1]{$x_1$}
\psfrag{y}[c][c][1]{$x_2$}
\includegraphics[height=5.5cm]{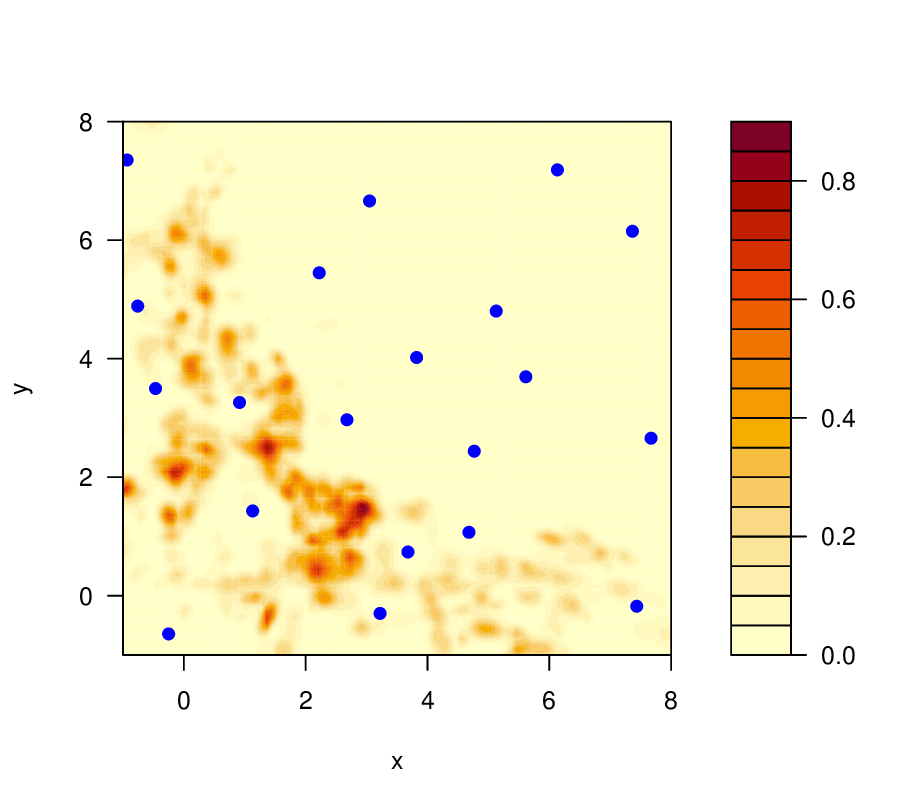}}
\caption{Graphical illustration of the reconstructions of $y$ (subfigure (a)) and $h^\star$ (subfigure (b)) from the mean of $Y_{N_0}$, denoted $\mu_{N_0}$, based on only $N_0=20$ noisy evaluations of the $y$ function. 
Subfigure (c) shows the empirical approximation of the standard deviation of the estimate of $h^\star$ from 100 independent runs of MCMC algorithms using 100 independent trajectories of $Y_{N_0}$.
In each figure, the blue dots correspond to the locations where the function $y$ was evaluated. In subfigures (a) and (b), the solid black lines correspond to the \emph{true} reference functions, and the red dotted lines correspond to the approximations from $\mu_{N_0}$.
}
\label{figCompMoyGPR}
}

\BFig{
\subfigure[Wasserstein distance]{
\psfrag{y}[c][c][1]{Wasserstein distance}
\psfrag{x}[c][c][0.7]{Number $N$ of noisy evaluations of $y$.}
\includegraphics[height=5.5cm]{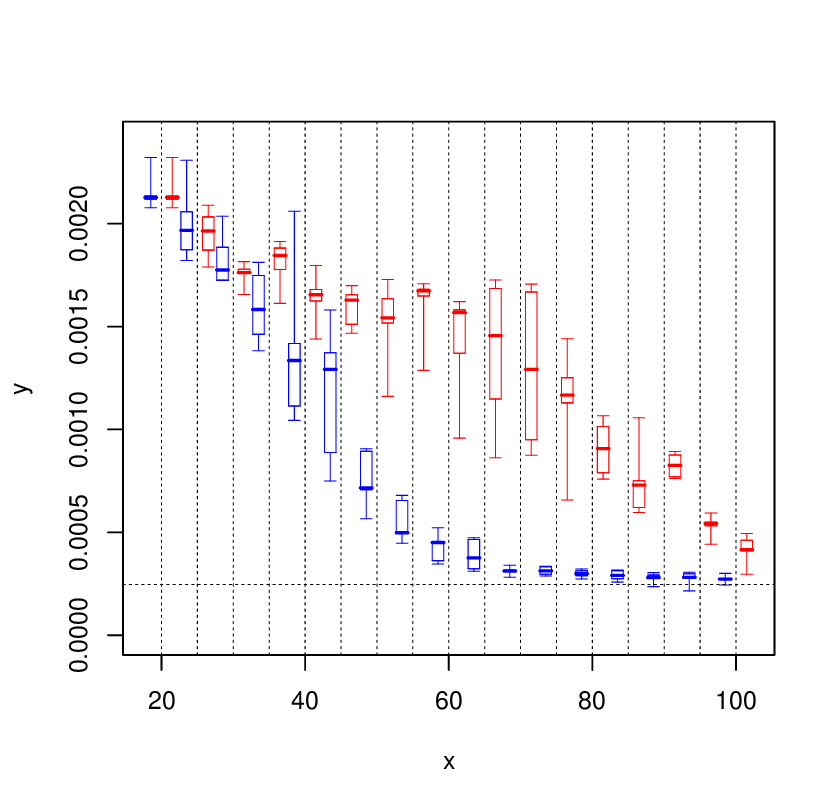}}
\subfigure[Variance]{
\psfrag{y}[c][c][1]{Integrated variance}
\psfrag{x}[c][c][0.7]{Number $N$ of noisy evaluations of $y$.}
\includegraphics[height=5.5cm]{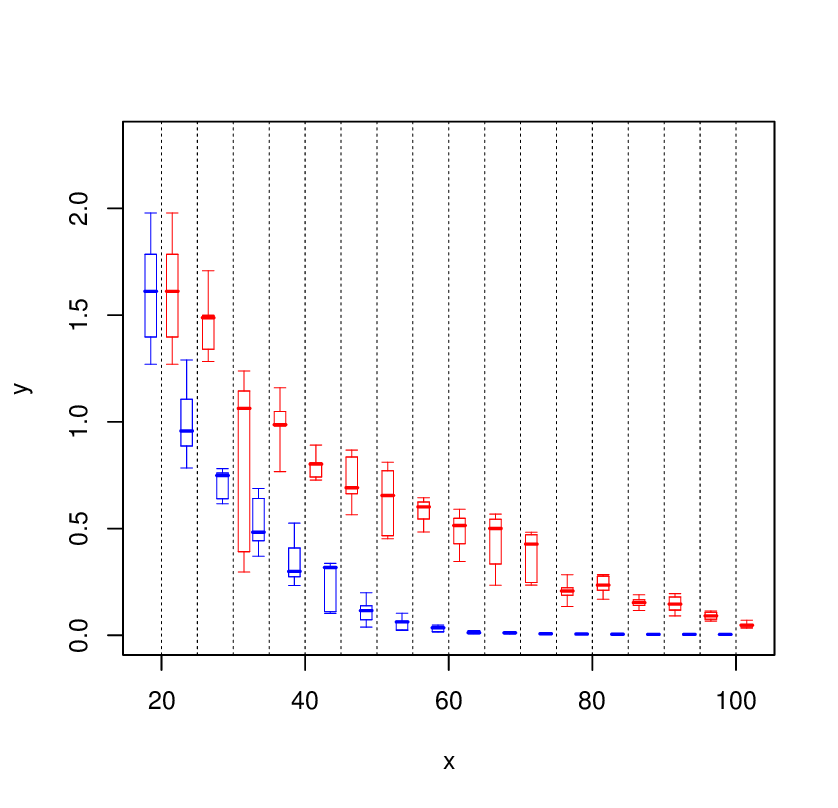}}
\caption{ Evolution of the reconstruction error (using the Wasserstein distance) and of the integrated variance with respect to the number $N$ of noisy evaluations of $y$. In (a), the horizontal line corresponds to the value of the reference error when the reconstruction of $h^\star$ relies on a huge number of calls to the true function $y$, while in (a) and (b), the vertical lines correspond to the specific values of $N$ where the errors are calculated. The blue boxplots, which are on the left of the vertical lines, are associated with the \emph{var-based} approach, and the red boxplots, which are on the right of the vertical lines, are associated with the \emph{space-filling} approach. }
\label{boxplot_analy}
}

This error convergence however requires knowledge of $h_\star$. If this reference function is not known, we can also look at the convergence of the variance of estimator $\widetilde{h}_{N,M,\widetilde{K}}$, as shown in Figure \ref{boxplot_analy}-(b). In this figure, the integrated variance is the integral over $\mathbb{X}$ of $\g{x}\mapsto \widetilde{V}_{N,M,\widetilde{K}}(\g{x})$, which is defined by Eq. (\ref{defVarh_tilde}). In the same manner as in Figure \ref{boxplot_analy}-(a), this figure shows that the results stabilize well in the \emph{var-based} approach, whereas convergence seems not to be  achieved yet for $N=100$ in the case of the \emph{space-filling} approach, hence the interest, once again, of the proposed sequential method.






\subsection{Application to the calibration of a high-speed train model}


The second example concerns the calibration of a longitudinal railway dynamics model (see \cite{nespoulous2022optimisation,nespoulous2024measurements} for more details about this simplified model and its parametrization) based on speed and energy consumption measurements. In this work, we limit ourselves to synthetic data, and we assume that we have access to $20$ measurements associated with the movement of the same train on $20$ different portions of French high speed lines. This set of measurements is divided in two: the first ten measurements are used to calibrate the train model, and serve as a training set; the last ten measurements are used to validate the calibration, and serve as a test set.
Two types of input must be specified for the numerical model. On the one hand, we denote by $\g{x}\in\mathbb{X}$ the vector of parameters characterizing the (dynamic and energetic) properties of the train that we are trying to estimate from on-track measurements. After an analysis of the model's sensitivities, only $d=9$ parameters are considered as sufficiently influential for this calibration phase: 
\BI{
\item $x_1$ is the total mass of the train and $x_2$ is the auxiliary power (or power deployed by the train for all activities other than traction and braking), 
\item $x_3$, $x_4$ and $x_5$ are three Davis coefficients used to define the aerodynamic friction force along the direction of the railway track in the form:

\Eqa{x_3+x_4\times (\dot{s}(t)-\dot{s}_{\text{wind}}(t))+x_5\times (\dot{s}(t)-\dot{s}_{\text{wind}(t)})^2,}

\noindent{}with $\dot{s}(t)$ and $\dot{s}_{\text{wind}}(t)$ the train and the wind speeds at time $t$,

\item $x_6,x_7,x_8$ and $x_9$ are four electrical efficiency coefficients, $x_6,x_7$ being associated with the conversion of the electrical power injected into the train into mechanical traction power, $x_8,x_9$ being associated with the conversion of the braking power into electrical power fed back into the catenary.
}

On the other hand, we denote by $z_\ell$ the driver command associated with the $\ell^{\text{th}}$ measurement, which is defined as a function indexed by the time and with values in $[-1,1]$. A command equal to $1$ indicates maximum traction, and a command equal to $-1$ indicates maximum braking. 
In a standard way, these functions are very close to piecewise constant functions (the driver does not continuously change the command, but only at certain times). Hence, for each measurement $\ell$, if we specify a certain value of $\g{x}$ in $\mathbb{X}=[0,1]^d$ and a certain driver command $z_\ell$, we can compute using the railway model the train speed $\dot{s}_\ell(t;\g{x},z_{\ell})$ and the consumed electric power $p_\ell(t;\g{x},z_{\ell})$ at any time $t\geq 0$. We can also introduce $e_{\ell}(t;\g{x},z_{\ell})$ as the energy consumed up to time $t$, so that :

\Eq{e_{\ell}(t;\g{x},z_{\ell})=\int_{0}^t p_\ell(t';\g{x},z_{\ell})dt'.}

From now on, we assume that there exists a true value for the train properties, which is chosen at random in $\mathbb{X}$ and noted $\g{x}^{\text{true}}$, as well as true driver controls for each of the 20 measurements considered.
These driver controls are chosen close to measured controls in terms of frequency of variation and change in control levels, and they are denoted by $z^{\text{true}}_{\ell}$. We thus denote by $p^{\text{true}}_\ell(t)=p_\ell(t;\g{x}^{\text{true}},z^{\text{true}}_{\ell})$ and $\dot{s}^{\text{true}}_\ell(t)=\dot{s}_\ell(t;\g{x}^{\text{true}},z^{\text{true}}_{\ell})$ the true power consumption and the true speed of the train on these different railway tracks, when we run the railway model with the true commands and the reference value for the train properties. We also suppose that we have access to noisy measurements of $p^{\text{true}}_\ell$ and $\dot{s}^{\text{true}}_\ell$ every $0.2s$ for each railway portion of the training set, which we gather in the vectors $\g{p}_\ell^{\text{mes}}$ and $\dot{\g{s}}_{\ell}^\text{mes}$ so that for each $1\leq \ell\leq 10$:

\BFig{
\hspace{-1cm}\subfigure[ \emph{Space-filling}, $N=40$ ]{
\psfrag{x}[c][c][1]{$x_1$}
\psfrag{y}[c][c][1]{$x_2$}
\includegraphics[height=4cm]{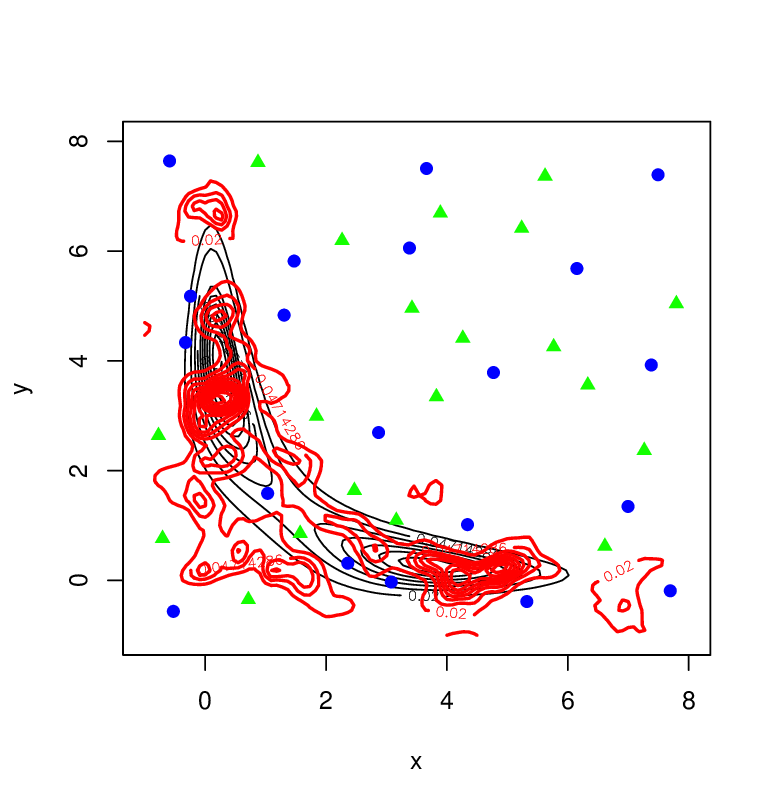}}
\subfigure[\emph{Space-filling}, $N=60$ ]{
\psfrag{x}[c][c][1]{$x_1$}
\psfrag{y}[c][c][1]{$x_2$}
\includegraphics[height=4cm]{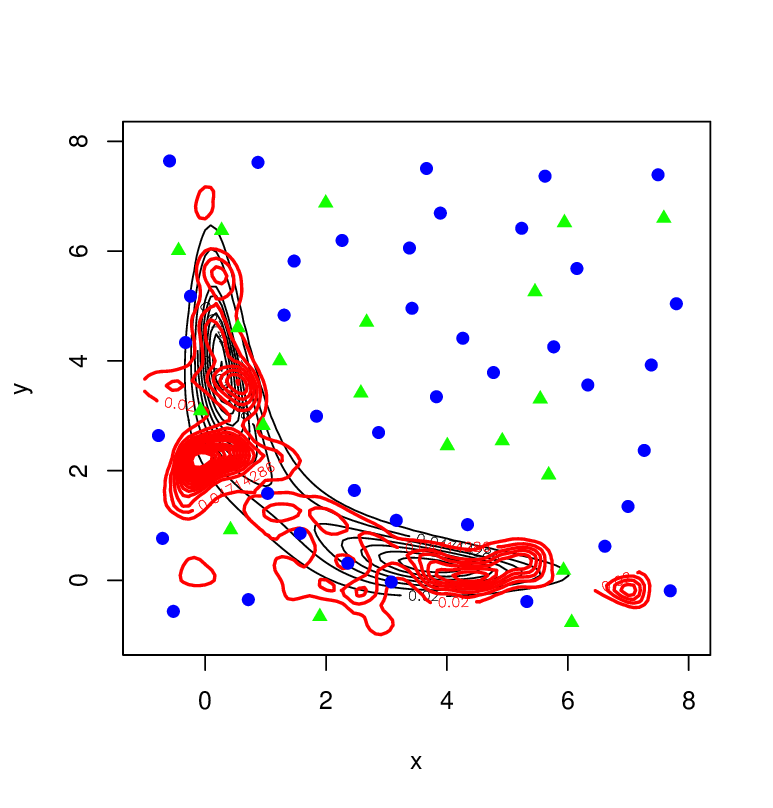}}
\subfigure[\emph{Space-filling}, $N=80$ ]{
\psfrag{x}[c][c][1]{$x_1$}
\psfrag{y}[c][c][1]{$x_2$}
\includegraphics[height=4cm]{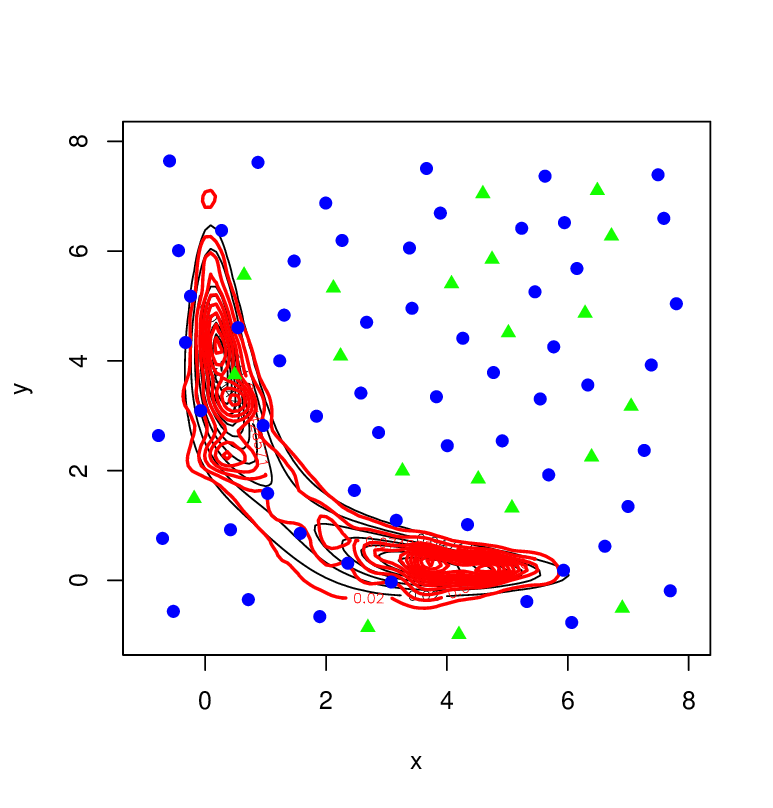}}
\subfigure[\emph{Space-filling}, $N=100$ ]{
\psfrag{x}[c][c][1]{$x_1$}
\psfrag{y}[c][c][1]{$x_2$}
\includegraphics[height=4cm]{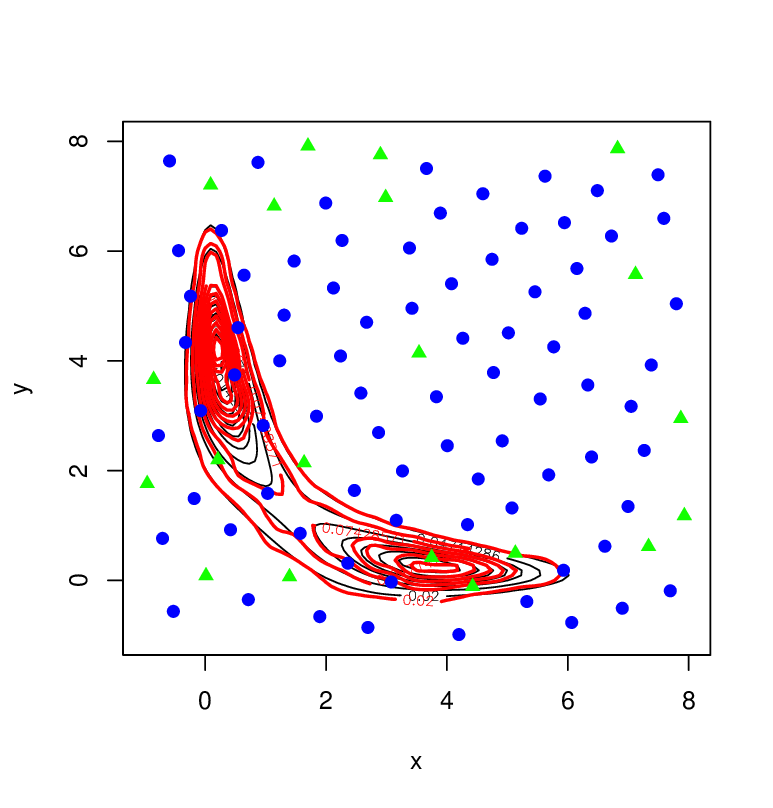}} \\
\vspace{-0.2cm}
\hspace{-1cm} \subfigure[ \emph{Var-based}, $N=40$ ]{
\psfrag{x}[c][c][1]{$x_1$}
\psfrag{y}[c][c][1]{$x_2$}
\includegraphics[height=4cm]{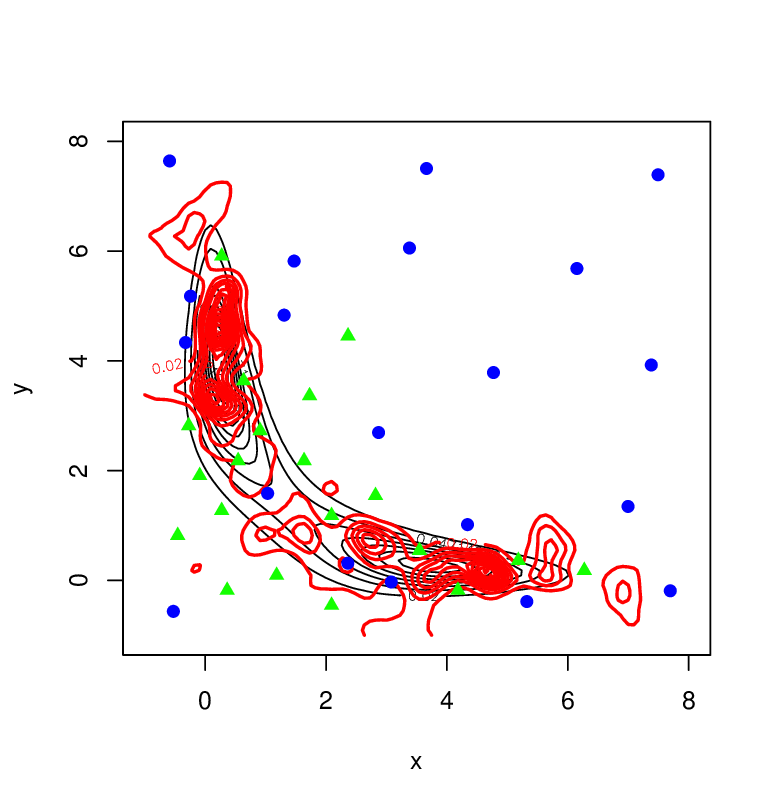}} 
\subfigure[ \emph{Var-based}, $N=60$]{
\psfrag{x}[c][c][1]{$x_1$}
\psfrag{y}[c][c][1]{$x_2$}
\includegraphics[height=4cm]{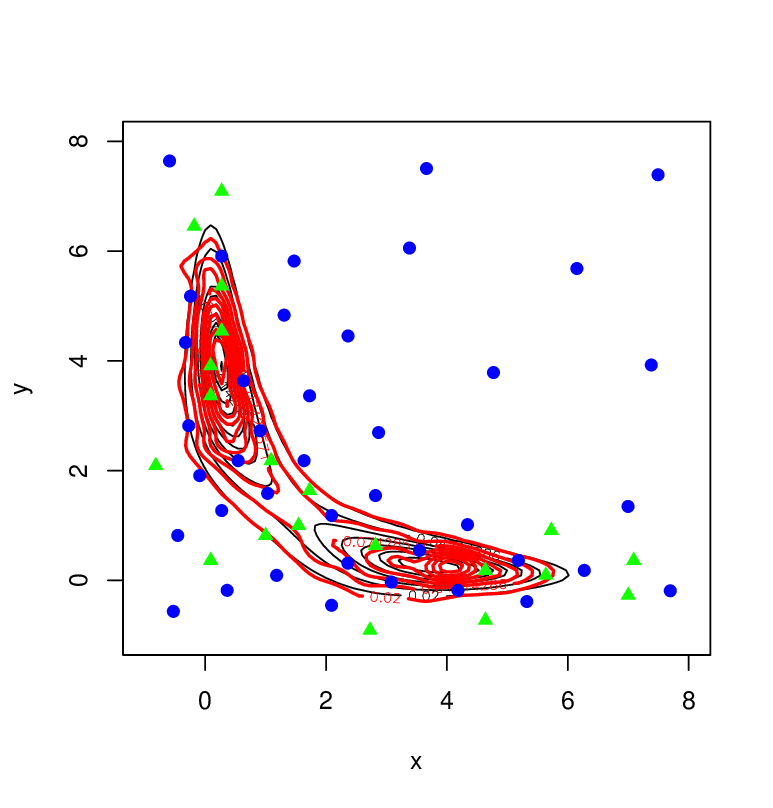}}
\subfigure[\emph{Var-based}, $N=80$ ]{
\psfrag{x}[c][c][1]{$x_1$}
\psfrag{y}[c][c][1]{$x_2$}
\includegraphics[height=4cm]{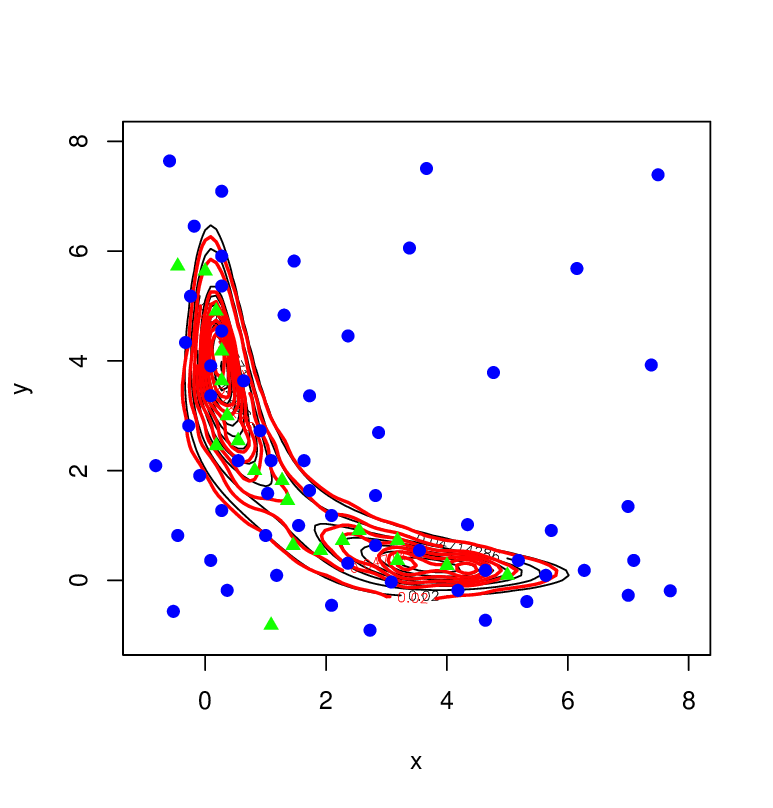}}
\subfigure[\emph{Var-based}, $N=100$ ]{
\psfrag{x}[c][c][1]{$x_1$}
\psfrag{y}[c][c][1]{$x_2$}
\includegraphics[height=4cm]{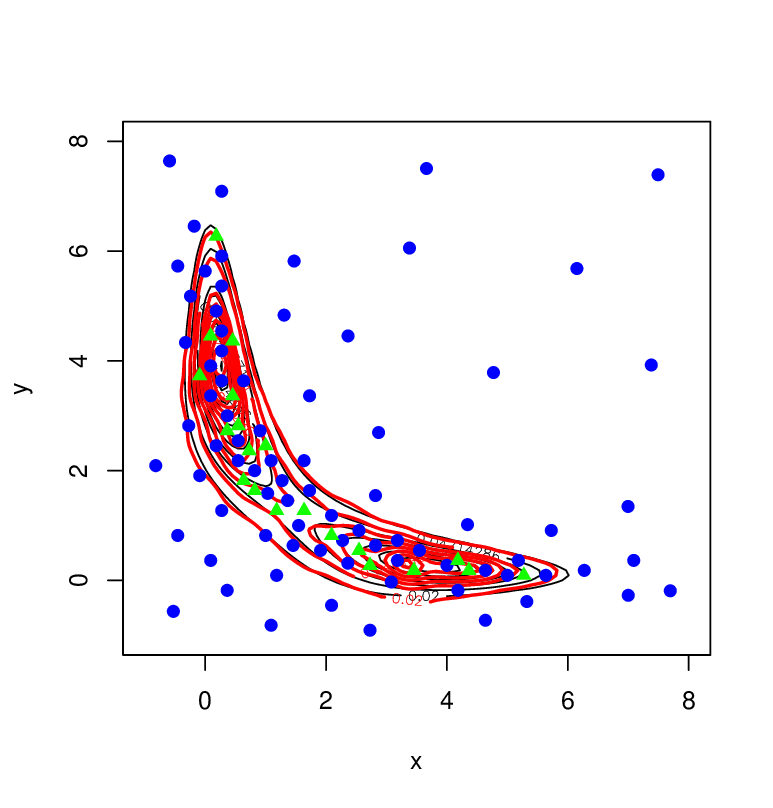}}
\caption{Graphical illustration of the reconstructions of $h^\star$ by the estimator $\widetilde{h}_{N,M,\widetilde{K}}$ defined by Eq. (\ref{estimatorFinal}) for different values of $N$ and two strategies (\emph{space-filling} and \emph{var-based}) for adding the new points.
In each figure, the blue dots and the green triangles correspond to the locations where the function $y$ was evaluated, the green triangles corresponding to the positions of the last 20 points added. The solid black lines correspond to the contourplots of $h^\star$, and the red lines correspond to the contourplots of $\widetilde{h}_{N,M,\widetilde{K}}$.
}
\label{figCompStrategieAdd}
}

\Eq{\g{p}_\ell^{\text{mes}}=\g{p}_\ell(\g{x}^{\text{true}},z_{\ell}^{\text{true}})+\g{\varepsilon}_\ell^{(p)}, \ \ \dot{\g{s}}_\ell^{\text{mes}}=\dot{\g{s}}_\ell(\g{x}^{\text{true}},z_{\ell}^{\text{true}})+\g{\varepsilon}_\ell^{(s)},\label{eqRefMes}}

\noindent{}where for each admissible (this notion of admissibility will be precised in the next subsection) values of $\g{x}$ and $z_{\ell}$, $\g{p}_\ell(\g{x},z_{\ell})=\PP{
p(0;\g{x},z_{\ell}),
p(0.2;\g{x},z_{\ell}),
\ldots,p(t^F_\ell;\g{x},z_{\ell})}$ and $\dot{\g{s}}_\ell(\g{x},z_{\ell})=\PP{
\dot{s}(0;\g{x},z_{\ell}),
\dot{s}(0.2;\g{x},z_{\ell}),
\ldots,\dot{s}(t^F_\ell;\g{x},z_{\ell})
}$ are the discretizations of $t\mapsto p(t;\g{x},z_{\ell})$ and $t\mapsto \dot{s}(t;\g{x},z_{\ell})$ at the $n_{\ell}=1+t^{F}_\ell/0.2$ measurement times, with $t^{F}_\ell$ the total duration of the train journey on the $\ell^{\text{th}}$ railway portion, and where $\g{\varepsilon}_\ell^{(p)}=\PP{
\varepsilon^{(p)}_1,
\ldots,
\varepsilon^{(p)}_{n_{\ell}}
} $ and $ \g{\varepsilon}^{(s)}_\ell=\PP{
 \varepsilon^{(s)}_1,
\ldots,
\varepsilon^{(s)}_{n_{\ell}} \\
} $ are two measurement errors.
To be consistent with the notations at the end of Section \ref{sec21}, we finally group all the measurement points in a vector $\g{m}$ of dimension $2\times \sum_{\ell=1}^{10}n_{\ell}$.




Following the developments of Section \ref{sec2}, the estimation of the train properties is carried out in a Bayesian perspective. 
In order to take account of their unknown nature, the vector gathering these parameters is modeled by a random vector $\g{X}$, with a prior PDF noted $f_{\g{X}}$ that is provided by experts in railway dynamics. For simplicity, we restrict ourselves in this work to the case where all these quantities have been normalized in a first phase, so that after normalization they are uniformly distributed on $[0,1]^d$:

\Eq{f_{\g{X}}(\g{x})= 1_{[0,1]^d}(\g{x}), \ \ \g{x}\in\R^d,}

\noindent{}where for any subset $A$ of $\R^d$, $1_A(\g{x})$ is equal to 1 if $\g{x}$ is in $A$ and to $0$ otherwise. Similarly, measurement errors are modeled by Gaussian quantities, and more precisely in this work as centered Gaussian vectors of respective covariance matrices $\sigma^2_s\g{I}_{n_{\ell}}$ and $\sigma^2_p\g{I}_{n_{\ell}}$, with $\sigma^2_s=0.1m/S$, $\sigma^2_p=10^2kW$, and $I_{d}$ the $(d\times d)$ identity matrix for any $d\geq 1$. The statistical properties of measurement errors are thus assumed to be known, and not to change from one measurement to the next. Let $\g{M}$ be the random vector gathering all the observations of speeds and powers consumed by the train under study. 
As explained at the end of Section \ref{sec21}, we can then use Bayes' theorem to deduce from Eq. (\ref{eqRefMes}) the \textit{a posteriori} PDF of $\g{X}$, which is denoted by $f_{\g{X}\vert \g{M}=\g{m}}$, and which is given by:

\Eq{f_{\g{X}\vert \g{M}=\g{m}}(\g{x}) \ \propto \ f_{\g{M}\vert \g{X}=\g{x}}(\g{m})f_{\g{X}}(\g{x}), \ \ \g{x}\in\R^d,}

\noindent{}where $\g{x}\mapsto f_{\g{M}\vert \g{X}=\g{x}}(\g{m})$ is the likelihood function. If the driver commands $z_\ell$ were perfectly known, this likelihood function could then be written as:

\Eq{f_{\g{M}\vert \g{X}=\g{x}}(\g{m}) \ \propto \
 \exp\PP{-\Ac{\sum_{\ell=1}^{10}\frac{\NOR{\g{p}_\ell(\g{x},z^{\text{true}}_{\ell})-\g{p}_{\ell}^{\text{mes}}}^2}{2\sigma^2_p}+
\frac{\NOR{\dot{\g{s}}_\ell(\g{x},z^{\text{true}}_{\ell})-\dot{\g{s}}_{\ell}^{\text{mes}}}^2}{2\sigma^2_s}}
}.\label{truelikely}}

In practice, however, these controls are often imperfectly known. It is typically the case when only a temporal discretization of these functions is available. For example, in this work, it is assumed that the driver controls can vary at any time, but that they are only recorded only every second. Hence, for each driver control $z_\ell^{\text{true}}$ that has been measured at the $Q$ instants $t_1\leq t_2\leq \ldots \leq t_Q$, if we note $\g{z}^{\ell,\text{disc}}:=(z^{\text{true}}_\ell(t_1),\ldots,z^{\text{true}}_\ell(t_Q))$ the discretization of $z^{\text{true}}_\ell$, we can introduce $\mathcal{Z}(\g{z}^{\ell,\text{disc}})$ as the set of commands that are consistent with $\g{z}^{\ell,\text{disc}}$, i.e. the set of piecewise constant functions $z$ with values in $[-1,1]$ such that for each $1\leq q\leq Q$ $z(t_q)=z_q^{\ell,\text{disc}}$. Note that the amplitude variations of the functions in $\mathcal{Z}(\g{z}^{\ell,\text{disc}})$ can take place at any time between two successive measurement points of $z^{\text{true}}_\ell$. As the true driver controls are assumed to be not perfectly known, they are modeled by random processes, which we assume to be independent statistically from one measurement to another, and uniformly distributed over $\mathcal{Z}(\g{z}^{\ell,\text{disc}})$, in the sense that the changes in control amplitude between two measurement times are assumed uniformly distributed over the interval separating these two measurements. For each $1\leq \ell\leq 10$, the random process associated with  $\mathcal{Z}(\g{z}^{\ell,\text{disc}})$ that characterizes the set of possible driver controls for the $\ell^{\text{th}}$ measurement is denoted by $Z_{\ell}$.  As explained in Section \ref{sec21}, the likelihood function becomes in that case

\Eq{f_{\g{M}\vert \g{X}=\g{x}}(\g{m}) \ \propto \
 \mathbb{E}\Cr{ \exp\PP{-\Ac{\sum_{\ell=1}^{10}\frac{\NOR{\g{p}_\ell(\g{x},Z_{\ell})-\g{p}_{\ell}^{\text{mes}}}^2}{2\sigma^2_p}+
\frac{\NOR{\dot{\g{s}}_\ell(\g{x},Z_{\ell})-\dot{\g{s}}_{\ell}^{\text{mes}}}^2}{2\sigma^2_s}}
} }
,\label{likelyhoodExpect}}

\noindent{}where the expectation is taken over the random driver controls $Z_{\ell}$. Note that according to this formalism, $z^{\text{true}}_{\ell}$ is a particular realization of $Z_{\ell}$.
By noting $\g{Z}$ the list gathering the $10$ random processes $Z_1,\ldots,Z_{10}$ associated with the uncertain driver controls, and by introducing the function $g$ so that

\Eq{g(\g{x},\g{Z})=f_{\g{X}}(\g{x})\times \exp\PP{-\Ac{\sum_{\ell=1}^{10}\frac{\NOR{\g{p}_\ell(\g{x},Z_{\ell})-\g{p}_{\ell}^{\text{mes}}}^2}{2\sigma^2_p}+
\frac{\NOR{\dot{\g{s}}_\ell(\g{x},Z_{\ell})-\dot{\g{s}}_{\ell}^{\text{mes}}}^2}{2\sigma^2_s}}
},}

\noindent{}we see that the posterior PDF $f_{\g{X}\vert \g{M}=\g{m}}$ of $\g{X}$, which we are trying to approximate, takes the same form as function $h^\star$ in Eq. (\ref{pb_mini}), and we find back the framework of this work:

\Eq{f_{\g{X}\vert \g{M}=\g{m}}(\g{x}) = c\times \mathbb{E}_{\g{Z}}\Cr{g(\g{x},\g{Z}) }, \ \ \ \ c^{-1}={\int_{\mathbb{X}}\mathbb{E}_{\g{Z}}\Cr{g(\g{x}',\g{Z})}d\g{x}'}.}


\BFig{
\subfigure[Train speed]{
\psfrag{x}[c][c][1]{$t$ (s)}
\psfrag{y}[c][c][1]{$\dot{s}_{\ell}$ (m/s)}
\psfrag{titre}[c][c][1]{}
\includegraphics[height=5.5cm]{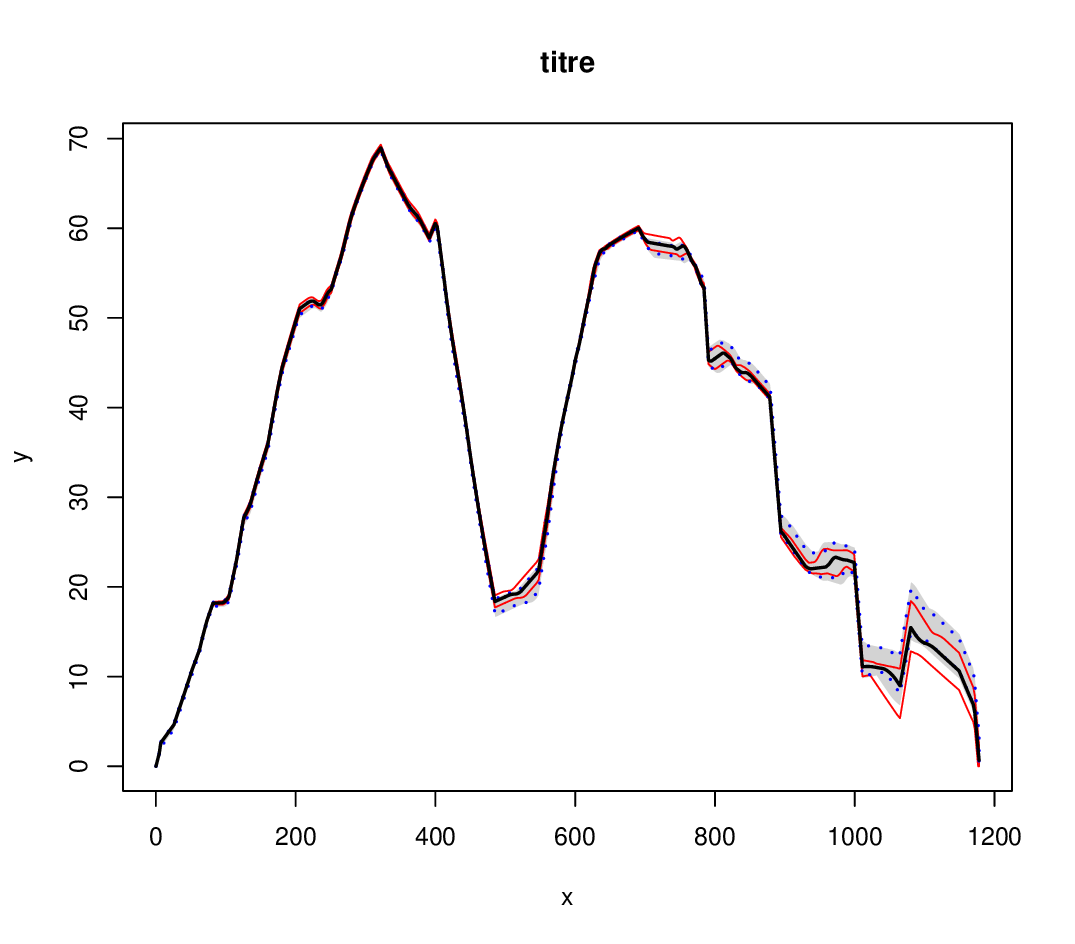}}
\subfigure[Electrical consumption]{
\psfrag{x}[c][c][1]{$t$ (s)}
\psfrag{y}[c][c][1]{$e_{\ell}$ (J)}
\psfrag{titre}[c][c][1]{}
\includegraphics[height=5.5cm]{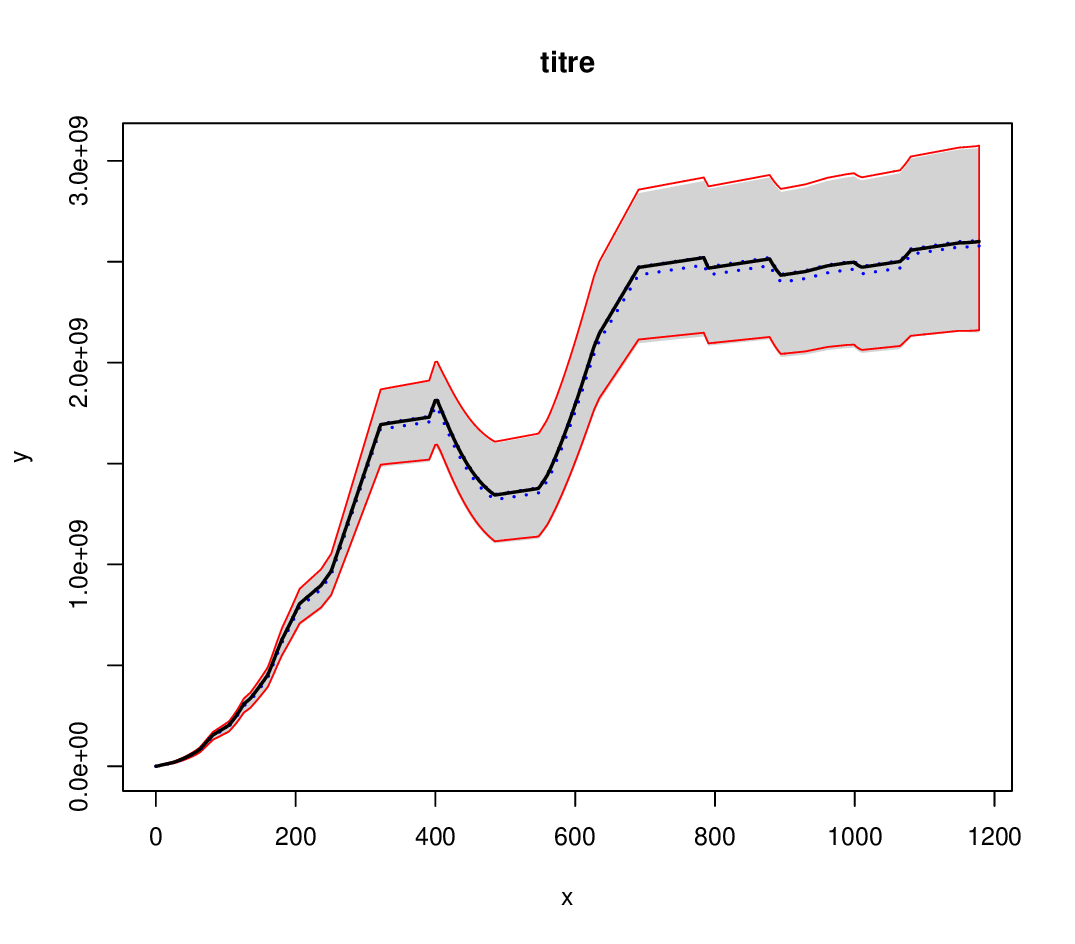}}
\subfigure[Driver controls]{
\psfrag{x}[c][c][1]{$t$ (s)}
\psfrag{y}[c][c][1]{$z_{\ell}$}
\psfrag{titre}[c][c][1]{}
\includegraphics[height=5.5cm]{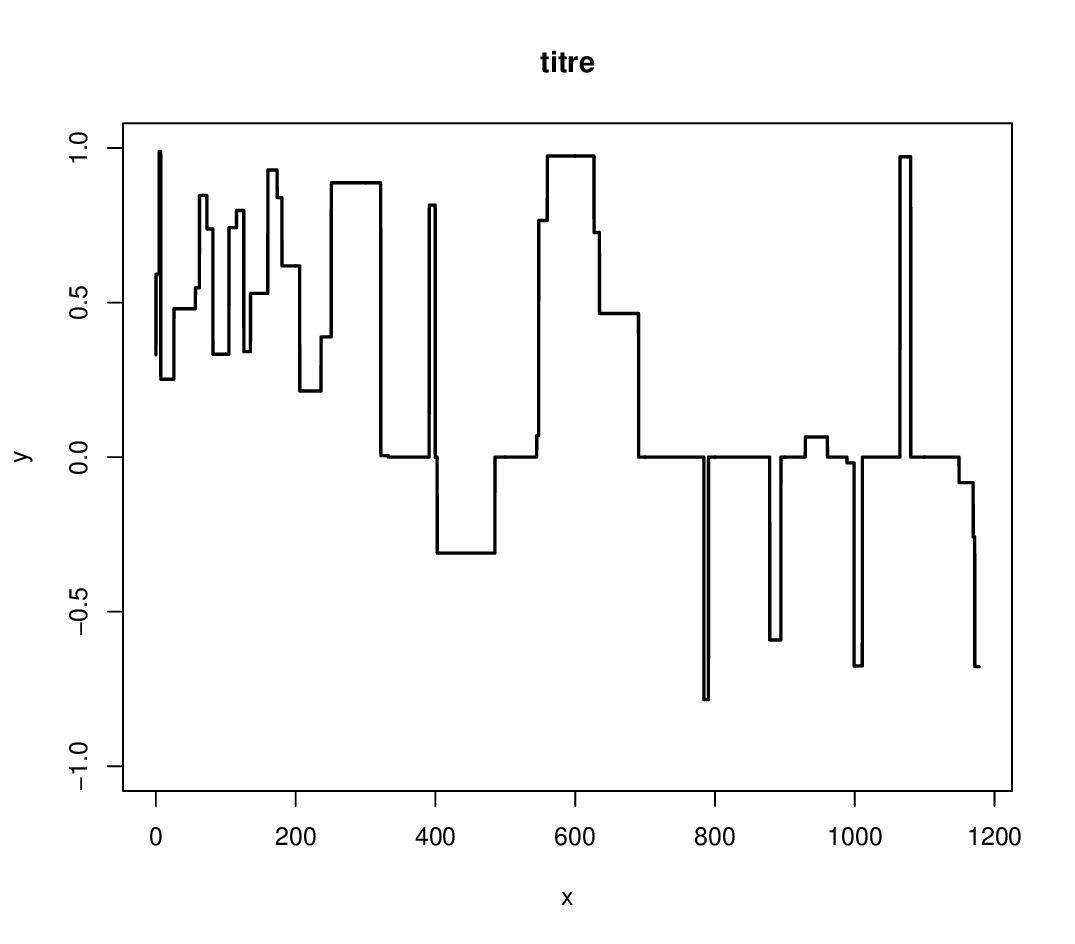}}
\caption{Evolutions of speed, consumed energy, and driver control as a function of time for the measurement $\ell=1$. The curves in solid thick black line correspond to the reference profiles, i.e. associated with the true driver control and the true model parameters. The gray area indicates the variation ranges that can be expected when simulating the dynamic and energetic behaviors of the train by varying the model parameters and the commands at the same time. The thin red lines indicate the variation ranges that can be expected by varying only the model parameters (and therefore considering the true driver command). The dotted blue lines indicate the variation ranges that can be expected by varying only the driver commands (and therefore considering the true model parameters).}
\label{fig_var_commandes}
}

Unsurprisingly, the train behavior is very sensitive to the uncertainties in the driver controls. As an illustration, Figure \ref{fig_var_commandes} compares, for one of the 10 measurements considered in the training set, the true speed and energy profiles with different speed and energy profiles when varying the properties $\g{x}$ of the train in $\mathbb{X}$ and the driver controls in $\mathcal{Z}(\g{z}^{\ell,\text{disc}})$. Note however in these graphs that the impact of changing the train or driver controls is of relatively the same order of magnitude when the train speed is considered, whereas the energy profile is mainly impacted by the train energy parameters. It can therefore be expected that the influence of the lack of knowledge with the driver controls will be all the more marked for the estimation of the dynamic parameters $(x_1^{\text{true}},x_3^{\text{true}},x_4^{\text{true}},x_5^{\text{true}})$ than for energetic parameters $(x_2^{\text{true}},x_6^{\text{true}},x_7^{\text{true}},x_8^{\text{true}},x_9^{\text{true}})$. 

\medskip

In order to assess the impact of this partial lack of knowledge of the driver controls, we introduce for any $1 \leq \ell \leq 10$ the driver controls $z^{\text{det}}_\ell$ as the elements of $\mathcal{Z}(\g{z}^{\ell,\text{disc}})$ such that the control changes always take place in the middle of the discretization steps.
These commands are thus constructed to be reasonable approximations of $z_\ell^{\text{true}}$, knowing that the chances of them perfectly matching the true commands are almost zero. Neglecting the uncertainty about the driver controls can then result in choosing a likelihood function proportional to 

\Eq{
 \exp\PP{-\Ac{\sum_{\ell=1}^{10}\frac{\NOR{\g{p}_\ell(\g{x},z^{\text{det}}_{\ell})-\g{p}_{\ell}^{\text{mes}}}^2}{2\sigma^2_p}+
\frac{\NOR{\dot{\g{s}}_\ell(\g{x},z^{\text{det}}_{\ell})-\dot{\g{s}}_{\ell}^{\text{mes}}}^2}{2\sigma^2_s}}
},}

\noindent{}instead of the one of Eq. (\ref{truelikely}). We can then couple these two likelihood functions (the first one associated with the true driver controls $z_\ell^{\text{true}}$ and the second one with the approximated ones $z_\ell^{\text{det}}$) to estimate the \textit{a posteriori} PDF of $\g{X}$ using an MCMC-type sampling approach based on evaluations of the direct railway code (without approximation). Let $\widehat{\g{x}}^{\text{true}}_1,\ldots,\widehat{\g{x}}^{\text{true}}_{\widetilde{K}}$ and $\widehat{\g{x}}^{\text{det}}_1,\ldots,\widehat{\g{x}}^{\text{det}}_{\widetilde{K}}$ be $\widetilde{K}=100$ points resulting from this MCMC approach, which can be considered i.i.d. according to the target probability distribution. Each time, around $K=10^5$ evaluations of the direct code on each of the 10 measurements of the training set were necessary to obtain these $\widetilde{K}$ points, while checking the convergence of the MCMC algorithm. In parallel, we can apply the approach detailed in Section \ref{sec2} to exploit the likelihood defined by Eq. (\ref{likelyhoodExpect}) at reasonable numerical cost. To do this, we replace, for each value of $\g{x}$ tested, the expectations associated with the uncertain driver controls on each railway track by a Monte Carlo estimator based on $R=100$ i.i.d. realizations of $Z_\ell$. We then obtain a noisy version of the expectation-based likelihood function, then of its logarithm. The logarithm of the posterior PDF of $\g{X}$ is then modeled by a Gaussian process, which is conditioned by a series of noisy evaluations of the logarithm of this likelihood function. The noise on these evaluations is assumed to be centered, and we propose to evaluate their variances using a bootstrap approach \cite{Bootstrap}. In this case, the variance is indeed heteroscedastic, changing from one log-likelihood evaluation to another. We can then couple this GPR approximation with a MALA-type MCMC algorithm as proposed in Algorithm \ref{algoMALA_GPR} to obtain ${\widetilde{K}}$ points that can be considered as i.i.d. and resulting from the same approximation of the \textit{a posteriori} distribution of $\g{X}$. And the more points we add for the GPR approximation, the more we expect these points to come from the "true" \textit{a posteriori} PDF, which is of course unknown for this application. In the following, we distinguish two cases: (1) the first one where the GPR approximation is based on $N_0=900$ noisy evaluations of the \textit{a posteriori} PDF in $N_0$ points uniformly chosen in $\mathbb{X}$ (requiring $N_0\times R = 9\times 10^4$ evaluations of the railway code for each of the $10$ track sections considered in the training set); (2) the second where the GPR approximation is based on the same $N_0$ first evaluations, but this time supplemented by $N-N_0=1500$ additional evaluations added in batches of 30 following enrichment procedure summarized in Section \ref{sec23}. We then note $\widehat{\g{x}}^{\text{unif}}_1,\ldots,\widehat{\g{x}}^{\text{unif}}_{\widetilde{K}}$ and $\widehat{\g{x}}^{\text{enriched}}_1,\ldots,\widehat{\g{x}}^{\text{enriched}}_{\widetilde{K}}$ the ${\widetilde{K}}$ points associated with these two cases respectively.
For comparison, we also note $\widehat{\g{x}}^{\text{prior}}_1,\ldots,\widehat{\g{x}}^{\text{prior}}_{\widetilde{K}}$ ${\widetilde{K}}$ i.i.d. realizations from the \textit{a priori }PDF of $\g{X}$. Two types of indicators can then be proposed to assess the relevance of the results. First, distinguishing between parameters associated with dynamics and those associated with energy consumption, for any set of ${\widetilde{K}}$ points $\widehat{\g{x}}_1,...\widehat{\g{x}}_{\widetilde{K}}$, we can approximate the mean square error (MSE) associated with the estimation of $\g{x}^{\text{true}}$, such that $\text{MSE}\approx \text{MSE}_v+\text{MSE}_p$, with


\Eq{\text{MSE}_v=\frac{1}{{\widetilde{K}}}\sum_{k=1}^{\widetilde{K}} \sum_{j\in\Ac{1,3,4,5}}\PP{\PP{\widehat{\g{x}}_k}_j-{x}_j^{\text{true}}}^2, \  \text{MSE}_p=\frac{1}{{\widetilde{K}}}\sum_{k=1}^{\widetilde{K}} \sum_{j\in\Ac{2,6,7,8,9}}\PP{\PP{\widehat{\g{x}}_k}_j-{x}_j^{\text{true}}}^2.}

Such an estimation error does not, however, distinguish between a highly influential parameter and one with little influence on the railway model (whether for the dynamic or the energy part). In order to integrate this dimension of model sensitivity, we can also calculate power prediction errors $\epsilon_{p,\ell}^2$ and speed prediction errors $\epsilon_{s,\ell}^2$ using the testing set measurements and the associated true driver controls, i.e. by choosing $\ell$ in $\Ac{11,\ldots,20}$, so that:

\Eq{
\epsilon_{s,\ell}^2 = \sum_{k=1}^{\widetilde{K}} \frac{\int_{0}^{t_{\ell}^F} \PP{\dot{s}(t;\widehat{\g{x}}_k,z^{\text{true}}_{\ell})-\dot{s}_\ell^{\text{true}}(t)}^2dt}{{\widetilde{K}}\times \int_{0}^{t_{\ell}^F} {\dot{s}_\ell^{\text{true}}(t)}^2dt}, \ \ 
\epsilon_{p,\ell}^2 = \sum_{k=1}^{\widetilde{K}} \frac{\int_{0}^{t_{\ell}^F} \PP{p(t;\widehat{\g{x}}_k,z^{\text{true}}_{\ell})-p_\ell^{\text{true}}(t)}^2dt}{{\widetilde{K}}\times \int_{0}^{t_{\ell}^F} p_\ell^{\text{true}}(t)^2dt}. }

The MSE errors are thus compared for the 5 configurations considered in Table \ref{tablefctTest}, while the prediction errors for the 10 test portions are shown as boxplots in Figure \ref{fig_var_commandes}. For all these indicators, we can see that the knowledge of driver controls is a definite advantage when estimating the train parameters, as the errors in this case are much smaller, both in terms of MSE and prediction errors. The prior case also gives us a reference point, by defining errors that are not based on any measurement. If we focus on the results associated with the points $\{\widehat{\g{x}}^{\text{det}}_k \}_{1\leq k\leq K}$, we can also quantify the negative effect of only knowing the driver command jumps every one second, without this being included in the definition of the likelihood function. In this case, the error indicators are indeed slightly better than the \textit{a priori} case, but multiplied by around $100$ compared to the reference. By integrating this uncertainty into the driver controls and adapting the likelihood function, we therefore see that it is possible to significantly reduce these errors, by a factor of $2$ for the MSE error, and by a factor of $10$ for the prediction errors. For this performance improvement, it is worth noting the vital importance of enrichment, as the case where we limit ourselves to a relatively small number of points uniformly distributed in $\mathbb{X}$ is clearly not sufficient, with errors close to the uniform case.

\begin{table}
\begin{center}
\begin{tabular}{c|ccccc}
Samples considered & $\widehat{\g{x}}_k^{\text{true}}$ & $\widehat{\g{x}}_k^{\text{prior}}$ &  $\widehat{\g{x}}_k^{\text{det}}$ & $\widehat{\g{x}}_k^{\text{unif}}$ & $\widehat{\g{x}}_k^{\text{enriched}}$ \\
\hline $\text{MSE}_v$ & 0.012 & 0.442 & 0.372 & 0.584 & 0.287   \\
$\text{MSE}_p$ & 0.001 & 0.720 & 0.545 & 1.100 & 0.304  \\
\hline $\text{MSE}$ & 0.013 & 1.162 & 0.917 & 1.684 & 0.592
\end{tabular}
\end{center}
\caption{Comparison of the MSE errors associated with the estimation of the train properties $\g{x}^{\text{true}}$ for the five considered configurations.}
\label{tablefctTest}
\end{table}

\BFig{
\subfigure[ Error on speed ]{
\psfrag{x}[c][c][1]{$t$ (s)}
\psfrag{y}[c][c][1]{$\epsilon_{s,\ell}^2$}
\psfrag{titre}[c][c][1]{}
\psfrag{a}[c][c][0.7]{$\widehat{\g{x}}_k^{\text{true}}$}
\psfrag{b}[c][c][0.7]{$\widehat{\g{x}}_k^{\text{prior}}$}
\psfrag{c}[c][c][0.7]{$\widehat{\g{x}}_k^{\text{det}}$}
\psfrag{d}[c][c][0.7]{$\widehat{\g{x}}_k^{\text{unif}}$}
\psfrag{e}[c][c][0.7]{$\widehat{\g{x}}_k^{\text{enriched}}$}
\includegraphics[height=5.5cm]{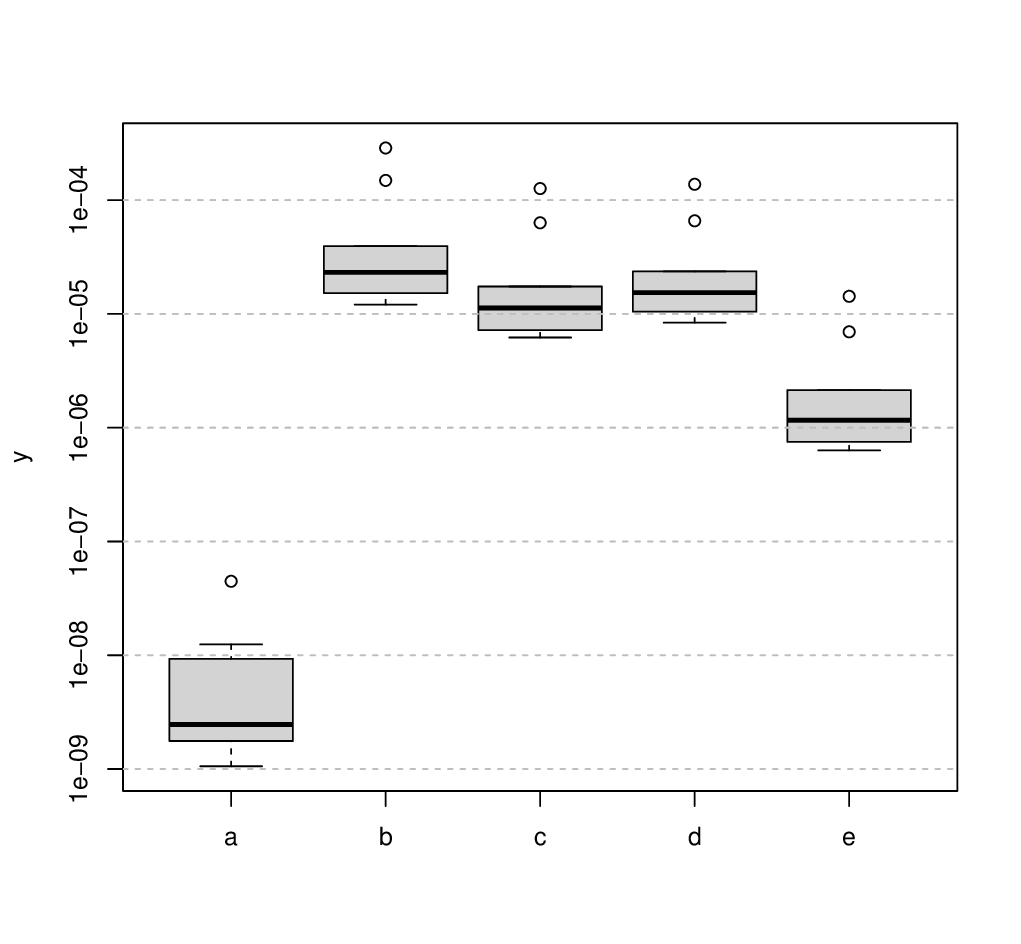}}
\subfigure[Error on electrical power]{
\psfrag{x}[c][c][1]{$t$ (s)}
\psfrag{y}[c][c][1]{$\epsilon_{p,\ell}^2$}
\psfrag{titre}[c][c][1]{}
\psfrag{a}[c][c][0.7]{$\widehat{\g{x}}_k^{\text{true}}$}
\psfrag{b}[c][c][0.7]{$\widehat{\g{x}}_k^{\text{prior}}$}
\psfrag{c}[c][c][0.7]{$\widehat{\g{x}}_k^{\text{det}}$}
\psfrag{d}[c][c][0.7]{$\widehat{\g{x}}_k^{\text{unif}}$}
\psfrag{e}[c][c][0.7]{$\widehat{\g{x}}_k^{\text{enriched}}$}
\includegraphics[height=5.5cm]{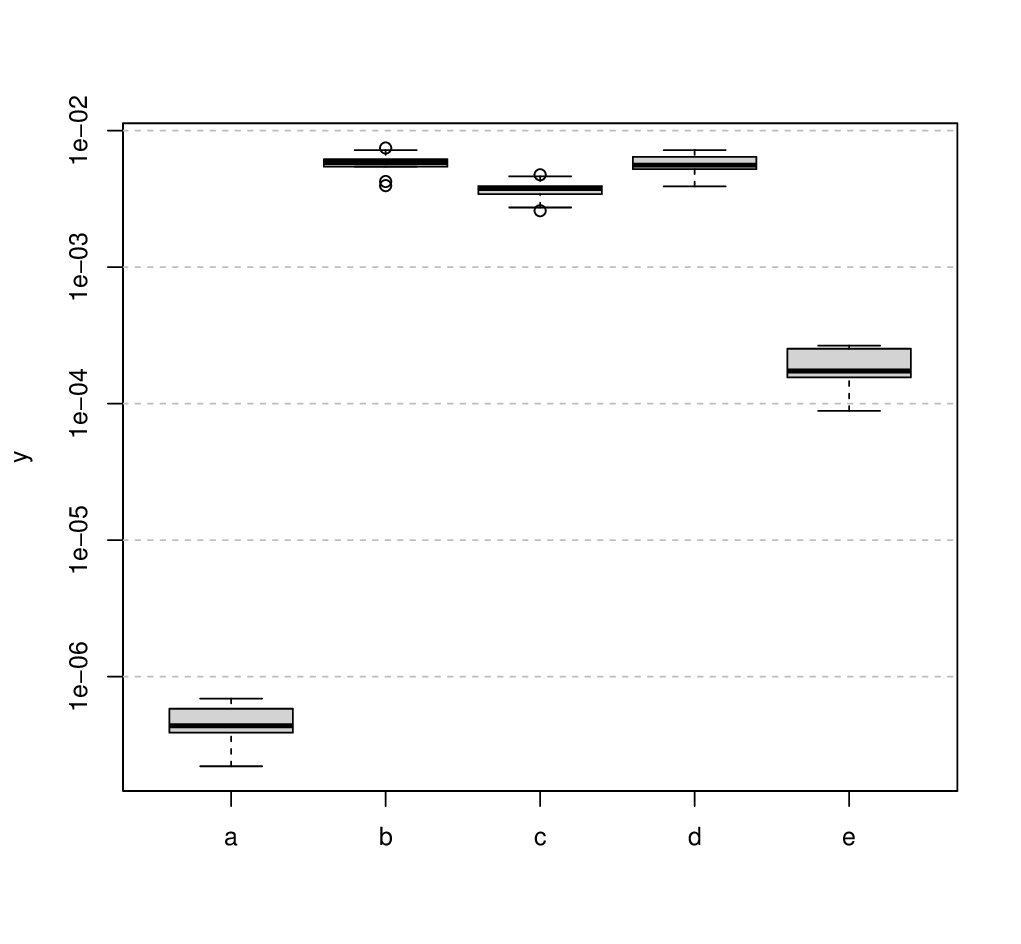}}
\caption{Comparison of the mean prediction errors for the five considered configurations.}
\label{fig_var_commandes}
}

\section{Conclusions and prospects}
\label{sec4}

This work introduces a methodology for the Bayesian calibration of computational models in situations where data are incomplete, which are conditions under which posterior distributions are often referred to as \emph{intractable}. The proposed approach combines Gaussian Process Regression (GPR) with the Metropolis-Adjusted Langevin Algorithm (MALA) to address the double challenge of expensive model evaluations and high-dimensional uncertainty propagation. A surrogate of the log-likelihood function is constructed using GPR, and simulation points are added sequentially, targeting regions of high posterior uncertainty to optimize the information gain while minimizing the computational cost. The resulting surrogate is then used within MALA to efficiently explore the posterior distribution while exploiting gradient information.
The method is evaluated on two representative cases: a synthetic analytical example where ground truth is known, and a real-world industrial scenario involving the estimation of mechanical and energy parameters of a high-speed train using on-board measurements. In both applications, the approach achieves accurate posterior inference with a limited number of costly simulations, demonstrating its interest.

This work also highlights several perspectives for improvement. The use of a logarithmic transformation of the likelihood function facilitated the surrogate modeling, but other transformations might better align the function with Gaussian process assumptions, and particularly in relation to error modeling. Indeed, here, the observation noise is modeled as Gaussian, but this assumption may not hold in all practical cases. Investigating transformations that induce Gaussianity in the residuals could improve robustness of the proposed method.
In addition, the current implementation uses a Gaussian process with a constant mean function. While this choice was effective in the tested scenarios, introducing richer mean structures, potentially informed by prior knowledge or data-driven trends, could further improve the surrogate prediction capacity, especially for transformed likelihoods with complex structure.


\section*{Acknowledgments}
The first author benefited from the support of the Chair Stress Test, Risk Management and Financial Steering, led by the French \'Ecole polytechnique and its foundation and sponsored by BNP Paribas.

\newpage

\bibliographystyle{plain}
\bibliography{bibliography}

\end{document}